\documentclass[twocolumn,longtable,times,tighten]{aastex631}

\usepackage[caption=false]{subfig}
\usepackage{mhchem}
\usepackage{float}
\DeclareCaptionLabelFormat{continued}{Continuation of #1~#2}
\begin{document}
\sloppy
\newcommand{\vdag}{(v)^\dagger}
\newcommand\aastex{AAS\TeX}
\newcommand\latex{La\TeX}
\newcommand{\mruncorr}{m$_{11}^\mathrm{R}$}
\newcommand{\mrcorr}{m$_{110}^\mathrm{R}$}
\def\absmag{H$_\mathrm{V}$}
\def\tss{$T_\mathrm{ss}$}
\def\rh{$r_\mathrm{h}$}
\def\espec{$\epsilon_\nu$}
\def\ebol{$\epsilon_\mathrm{bol}$}
\def\deq{$D_\mathrm{eq}$}
\def\geomalb{p$_\mathrm{V}$}
\def\qg{2001\,QG$_{298}$}
\def\deff{D$_\mathrm{eff}$}
\def\gcc{g\,cm$^{-3}$}
\def\tiunit{$\mathrm J\,m^{-2}\,s^{-1/2}K^{-1}$}

\received{\today}
\revised{\today}
\accepted{\today}
\submitjournal{ApJL}

\shorttitle{Prominent mid-infrared excess of Makemake}
\shortauthors{Kiss et al.}

\graphicspath{{./}{figures/}}

\title{Prominent mid-infrared excess of the dwarf planet (136472) Makemake \\
discovered by JWST/MIRI indicates ongoing activity}

\correspondingauthor{Csaba Kiss}
\email{kiss.csaba@csfk.org}


\author[0000-0002-8722-6875]{Csaba Kiss}
\affiliation{Konkoly Observatory, HUN-REN Research Centre for Astronomy and Earth Sciences, Konkoly Thege 15-17, 1121 Budapest, Hungary}
\affiliation{CSFK, MTA Centre of Excellence, Konkoly Thege 15-17, 1121, Budapest, Hungary}
\affiliation{ELTE E\"otv\"os Lor\'and University, Institute of Physics and Astronomy, P\'azm\'any P\'eter s\'et\'any 1/A, 1171 Budapest, Hungary}

\author[0000-0002-0717-0462]{Thomas G. M\"uller}
\affiliation{Max-Planck-Institut f\"ur extraterrestrische Physik, Giessenbachstr. 1, 85748 Garching, Germany}
\author[0000-0001-5531-1381]{Anik\'o Farkas-Tak\'acs}
\affiliation{Konkoly Observatory, HUN-REN Research Centre for Astronomy and Earth Sciences, Konkoly Thege 15-17, 1121 Budapest, Hungary}
\affiliation{CSFK, MTA Centre of Excellence, Konkoly Thege 15-17, 1121, Budapest, Hungary}
\author[0009-0001-9360-2670]{Attila Mo\'or}
\affiliation{Konkoly Observatory, HUN-REN Research Centre for Astronomy and Earth Sciences, Konkoly Thege 15-17, 1121 Budapest, Hungary}
\affiliation{CSFK, MTA Centre of Excellence, Konkoly Thege 15-17, 1121, Budapest, Hungary}
\author[0000-0001-8541-8550]{Silvia Protopapa}
\affiliation{{Southwest Research Institute, 1301 Walnut Street, Suite 400, Boulder, Colorado 80302, USA}}
\author[0000-0002-6722-0994]{Alex H. Parker}
\affiliation{SETI Institute, Mountain View, CA, USA}
\author[0000-0002-1123-983X]{Pablo Santos-Sanz}
\affiliation{Instituto de Astrof\'{\i}sica de Andaluc\'{\i}a (CSIC), Glorieta de la Astronom\'{\i}a s/n, 18008 Granada, Spain}
\author[0000-0002-8690-2413]{Jose Luis Ortiz}
\affiliation{Instituto de Astrof\'{\i}sica de Andaluc\'{\i}a (CSIC), Glorieta de la Astronom\'{\i}a s/n, 18008 Granada, Spain}
\author[0000-0002-6117-0164]{Bryan J. Holler}
\affiliation{Space Telescope Science Institute, 3700 San Martin Drive, Baltimore, MD 21218, USA}
\author[0000-0001-9665-8249]{Ian Wong}
\affiliation{NASA Goddard Space Flight Center, Greenbelt, MD 20771, USA}
\author[0000-0003-2434-5225]{John Stansberry}
\affiliation{Space Telescope Science Institute, 3700 San Martin Drive, Baltimore, MD 21218, USA}
\author[0000-0003-2132-7769]{Estela Fern\'andez-Valenzuela}
\affiliation{Florida Space Institute, University of Central Florida, 12354 Research Parkway, 32826 Orlando, USA}
\author[0000-0002-2161-4672]{Christopher R. Glein}
\affiliation{Space Science Division, Space Sector, Southwest Research Institute, 6220 Culebra Road, San Antonio, TX 78238-5166, USA}
\author[0000-0001-7168-1577]{Emmanuel Lellouch}
\affiliation{LESIA, Observatoire de Paris, PSL Research University, CNRS, Sorbonne Université, UPMC Univ. Paris 06, Univ. Paris Diderot, Sorbonne Paris Cité, 5 place Jules Janssen, 92195 Meudon, France}
\author[0000-0002-6184-7681]{Esa Vilenius}
\affiliation{Mullard Space Science Laboratory, University College London, UK}
\author[0000-0002-1663-0707]{Csilla E. Kalup}
\affiliation{Konkoly Observatory, HUN-REN Research Centre for Astronomy and Earth Sciences, 
Konkoly Thege 15-17, 1121 Budapest, Hungary}
\affiliation{CSFK, MTA Centre of Excellence, Konkoly Thege 15-17, 1121, Budapest, Hungary}
\affiliation{ELTE E\"otv\"os Lor\'and University, Institute of Physics and Astronomy, P\'azm\'any P\'eter s\'et\'any 1/A, 1171 Budapest, Hungary}
\author[0000-0001-5573-8190]{Zsolt Regály}
\affiliation{Konkoly Observatory, HUN-REN Research Centre for Astronomy and Earth Sciences, 
Konkoly Thege 15-17, 1121 Budapest, Hungary}
\affiliation{CSFK, MTA Centre of Excellence, Konkoly Thege 15-17, 1121, Budapest, Hungary}
\author[0000-0002-1698-605X]{R\'obert Szak\'ats}
\affiliation{Konkoly Observatory, HUN-REN Research Centre for Astronomy and Earth Sciences, 
Konkoly Thege 15-17, 1121 Budapest, Hungary}
\affiliation{CSFK, MTA Centre of Excellence, Konkoly Thege 15-17, 1121, Budapest, Hungary}

\author[0000-0002-1326-1686]{G\'abor Marton}
\affiliation{Konkoly Observatory, HUN-REN Research Centre for Astronomy and Earth Sciences, 
Konkoly Thege 15-17, 1121 Budapest, Hungary}
\affiliation{CSFK, MTA Centre of Excellence, Konkoly Thege 15-17, 1121, Budapest, Hungary}

\author[0000-0001-5449-2467]{Andr\'as P\'al}
\affiliation{Konkoly Observatory, HUN-REN Research Centre for Astronomy and Earth Sciences, 
Konkoly Thege 15-17, 1121 Budapest, Hungary}
\affiliation{CSFK, MTA Centre of Excellence, Konkoly Thege 15-17, 1121, Budapest, Hungary}
\affiliation{ELTE E\"otv\"os Lor\'and University, Institute of Physics and Astronomy, P\'azm\'any P\'eter s\'et\'any 1/A, 1171 Budapest, Hungary}

\author[0000-0002-0606-7930]{Gyula M. Szab\'o}
\affiliation{ELTE E\"otv\"os Lor\'and University, Gothard Astrophysical Observatory, Szombathely, Hungary}
\affiliation{MTA-ELTE Exoplanet Research Group, 9700 Szombathely, Szent Imre h. u. 112, Hungary}

\begin{abstract}
We report on the discovery of a very prominent mid-infrared (18-25\,$\mu$m) excess associated with the trans-Neptunian dwarf planet (136472) Makemake. The excess, detected by the MIRI instrument of the James Webb Space Telescope, along with previous measurements from the Spitzer and Herschel space telescopes, indicates the occurence of temperatures of $\sim$150\,K, much higher than what solid surfaces at Makemake's heliocentric distance could reach by solar irradiation. We identify two potential explanations: a continuously visible, currently active region, powered by subsurface upwelling and possibly cryovolcanic activity, covering $\leq$1\% of Makemake's surface, or an as yet undetected ring containing very small carbonaceous dust grains, which have not been seen before in trans-Neptunian or Centaur rings.  Both scenarios point to unprecedented phenomena among trans-Neptunian objects and could greatly impact our understanding of these distant worlds.

\end{abstract}


\section{Introduction \label{introduction}}

(136472) Makemake is one of the largest and brightest objects in the Kuiper belt. \citet{Ortiz2012} derived size and albedo for Makemake from occultation measurements and obtained an equivalent diameter of $\sim$1430\,km, an intermediate size between that of Pluto/Eris and Charon. They also found a bright surface with a geometric albedo of $p_V$\,$\approx$\,0.8. The surface is known to be predominantly covered by methane (CH$_4$) ice \citep{Brown2007,AC2020}, and by CH$_{4}$ irradiation products \citep{Brown2015}. 
\citet{Grundy2024} report that the D/H ratio in CH$_4$ ice observed on  Makemake is significantly lower than that detected in comet 67P/Churyumov-Gerasimenko, which is considered to be primordial.
However, it aligns closely with the ratios found in water in many comets and larger outer solar system objects. These similarities and differences prompted \cite{Glein2024} to suggest that the hydrogen atoms in CH$_{4}$ on Makemake originated from water, generated by geochemical processes at elevated temperatures in the deep interior.
While there are several features on the surface of trans-Neptunian objects indicating past cryovolcanism \citep[see][for a summary]{cryo}, ongoing activity has not been observed so far. 

Thermal emission measurements in the infrared (IR) are traditionally used to obtain the size and albedo of solar system bodies \citep{Muller2020}. In addition, they also put constraints on the object’s thermal properties and spin-axis orientations.
The integration of multi-wavelength and multi-technique data enhances the physical and thermal characterization of trans-Neptunian objects. This comprehensive approach allows for the inclusion of additional components, such as satellites and rings, and helps constrain properties that are otherwise inaccessible \citep[see e.g.][]{Lellouch2017,Muller2019,Kiss2024}. 

The thermal emission of Makemake was first measured by the Spitzer Space Telescope \citep{Stansberry2008} and then by the Herschel Space Observatory in the Science Demonstration Phase \citep{Lim2010}. To fit the observed flux densities, \citet{Lim2010} proposed a double terrain model. A dark / warm component was {necessary in addition to the general cold/high albedo surface} to account for the excess observed by the Spitzer/Multiband Imaging Photometer (MIPS) at 24\,$\mu$m.
It was also suggested that the dark terrain {might represent a satellite that was unknown at the time.}
\citet{Parker2016} detected a satellite using Hubble Space Telescope (HST) measurements, which was 7.80\,mag fainter than the primary. They proposed that this satellite might contribute to the 24-$\mu$m excess emission. By means of the Near-Earth Asteroid Thermal Model \citep[see e.g.][]{Lellouch2013}, they found that to match the observed flux densities, the satellite contribution required a beaming parameter value (which describes the deviation of the surface temperature from that of a smooth surface in instantaneous equilibrium) of $\eta$\,$\leq$\,0.4. This very low value is difficult to reconcile with the characteristics observed on real surfaces \citep{Spencer1990,Lellouch2013}. 
In the study by \citet{Lellouch2017}, ALMA band-6 (1.3\,mm) measurements of Makemake were used to {determine} the submillimetre emissivity. They obtained a relative emissivity of $\epsilon_r$\,$\approx$\,1, in contrast to the $\epsilon_r$\,$\approx$\,0.7 {value typically observed on the}  surfaces of most Centaurs and trans-Neptunian objects. 
In the same paper, the authors explored two separate scenarios to model the thermal emission, incorporating the Spitzer/MIPS 24\,$\mu$m measurement. One scenario involved a very dark satellite, while the other considered thermal emission from diffuse dust in instantaneous equilibrium with solar radiation. While technically both models could fit the observations to some acceptable level, they both had issues with the physical interpretation of the parameters obtained, and some open issues remained: 
1) {Is the mid-infrared thermal emission excess observed at 24\,$\mu$m during a single epoch a permanent feature?}
2) {Could further measurements of thermal emission provide additional constraints on the thermal properties?} 
3) {Is it possible to discern which of the proposed models—dark terrain, dark satellite, or diffuse dust—is best suited to explain the observations?} 

To answer these key questions we conducted a comprehensive investigation of the thermal emission of the Makemake system 
by integrating new measurements from the James Webb Space Telescope's Mid-Infrared Instrument (MIRI), unpublished Herschel/PACS observations, and a re-evaluation of previously published data.  Additionally, we included visible light curve data from the TESS and Gaia space telescopes to constrain the rotation period, and unpublished Spitzer/MIPS data at 24 and 70\,$\mu$m from a second epoch, covering a substantial portion of Makemake's rotation period, to obtain a partial thermal light curve and constrain rotational thermal emission variations.  

\section{Observational data and thermal emission modeling \label{sect:thermalmodelmain}}

The details of the observations and data reduction are presented in Sects.~\ref{sect:period} and \ref{sect:thermal} in the Appendix. 

Visible range light curve measurements can provide rotational properties, which are essential for interpreting thermal emission data. We used data from the TESS and Gaia space telescopes. The light curve measurements from  both TESS and Gaia confirm the previously established 11.4\,h single-peaked rotation period obtained by \cite{Hromakina2019}. The TESS double-peaked light curve (using a period of 22.8\,h) does not show a significant asymmetry between the light curves of the two half periods, i.e. we cannot confirm that the double-peaked 22.8\,h is Makemake's true rotation period. In the following we use P\,=\,11.4\,h as the default rotation period, but in some cases we also perform calculations using P\,=\,22.8\,h.

\medskip

The thermal emission measurements of Makemake cover the wavelength range from 18\,$\mu$m to 1.3\,mm, observed by Spitzer/MIPS, Herschel/PACS and SPIRE, and ALMA; the latest measurements were performed with JWST/MIRI imaging in the F1800W and F2550W bands at 18.0 and 25.5\,$\mu$m, respectively (see Tables \ref{table:allmeas} and \ref{table:all_flux} in the Appendix). These measurements are typically short snaphots, except a longer Spitzer/MIPS 24 and 70\,$\mu$m measurement sequence of 7.6\,h. 

Although the Spitzer/MIPS 24\,$\mu$m may indicate small brightness variation ($\sim$16\% peak-to-peak of the mean flux density level, see Sect.~\ref{sect:thermal}), both the 24 and 70\,$\mu$m long Spitzer/MIPS measurements are compatible with a constant light curve within $\lesssim$25\% of the mean flux density level, and are also compatible with the other 24\,$\mu$m Spitzer measurements within this uncertainty.
An obvious result from the thermal emission data is that the latest (January 2023) JWST/MIRI measurements confirm the previously seen high mid-infrared flux densities at $\sim$25\,$\mu$m, and also show very high values, especially in the F1800W band (166 and 356\,$\mu$Jy at 18 and 25.5\,$\mu$m, respectively). Thermal emission models of the Makemake system should be able to reproduce and explain the likely origin of these features. 

To model the thermal emission of Makemake we first used the Near-Earth Asteroid Thermal model \citep[{NEATM,}][]{Harris1998}. 
Although this is a simple, compressed parameter model, its usage is justified as we lack detailed information on Makemake's spin and shape properties, and we have only limited rotationally resolved thermal emission data. The NEATM concept uses the beaming parameter $\eta$ as a proxy for thermal effects related to thermal inertia, surface roughness, spin rate or subsolar latitude, used in more complicated thermophysical models \citep{Spencer1990,Lellouch2013}. 

The NEATM model considers the observing geometry via the heliocentric distance of the target ($r_h$), the observer distance ($\Delta$) and the phase angle ($\alpha$). We transformed all observed flux densities to a common observing geometry of $r_h$\,=\,52\,au, $\Delta$\,=\,52\,au, and $\alpha$\,=\,1\degr. We note, however, that the observing geometry changed only slightly between the different epochs, and the difference between the corrected and uncorrected flux densities are $\lesssim$4\%, which is smaller than the relative uncertainties in the measured in-band flux densities ($\gtrsim$10\%), and the additional $\sim$5\% absolute calibration errors of the detectors. 
Sub-mm and mm data (Herschel/SPIRE and ALMA band-6) have been sourced from \citet{Lellouch2017}. All other previously available infrared data (Spitzer/MIPS and Herschel/PACS) have been reevaluated using the latest versions of the respective pipelines (see Sect.~\ref{sect:thermal}). 
The 24\,$\mu$m thermal light curve shows only a small amplitude variation, which may even correspond to a constant light curve, and we have no reliable light curve information for the other wavelengths, however, multi-epoch observations show similar values. Therefore we used the weighted mean values per instrument/filter in our thermal emission modeling. These values were calculated from the data presented in Table~\ref{table:all_flux}.

\begin{figure*}[ht!]
    \centering
    \includegraphics[width=\textwidth]{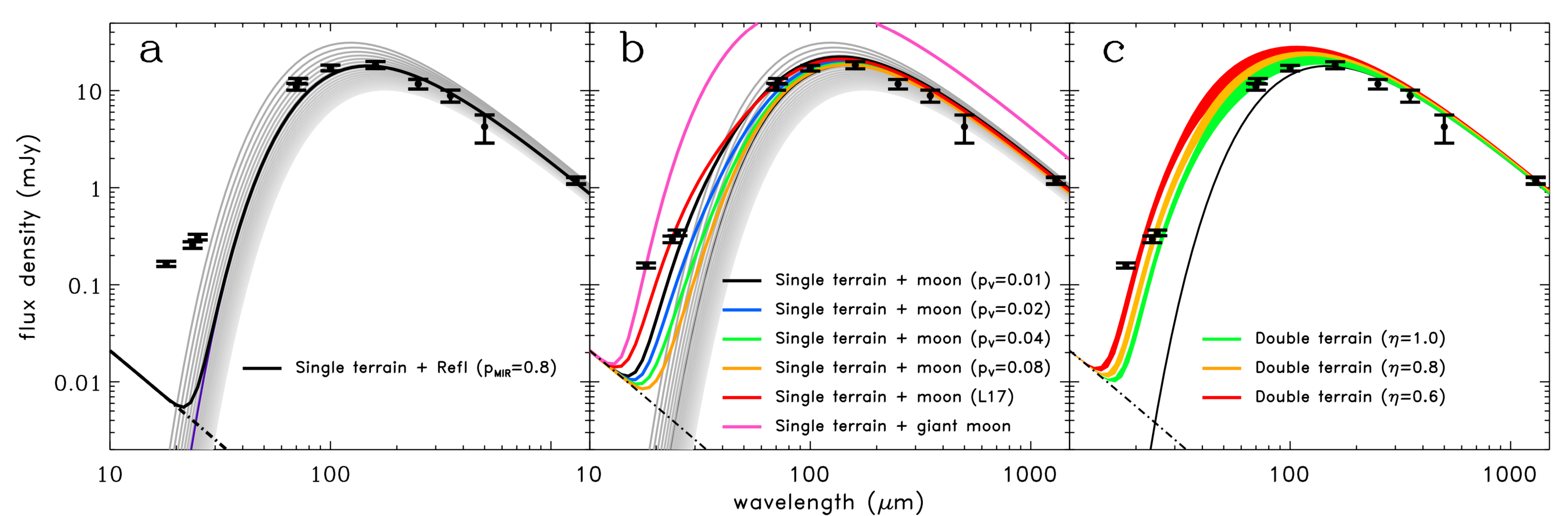}
    \caption{Spectral energy distribution of Makemake from NEATM modeling. 
    a) Assuming a single terrain. Gray curves correspond to Case 1 surfaces, assuming different $\eta$ beaming parameters. The black solid curve has $\eta$\,=\,1.2. The black dash-dotted curve corresponds to a reflected light contribution assuming a mid-infrared albedo of $p_{MIR}$\,=\,0.8 in all subfigures. 
    b) Case 2: Makemake's thermal emission model considering a single-terrain Makemake and a dark satellite. The curves with different colours correspond to satellites with $p_V$\,=\,0.01...0.08, the 'extreme' moon model used by \citet[][marked as L17]{Lellouch2017}, and the dark giant moon case, as indicated by the inserted text. 
    c) Makemake's thermal emission considering double-terrain models (Case 3). The bands with different colours cover the models using different bright/dark terrains (see Table~\ref{table:terrains}) and dark terrain locations, but using the same beaming parameters, as shown by the insert.  }
    \label{fig:neatm3}
\end{figure*}

We attempted to fit Makemake's thermal emission using three kind of model configurations (see details in Sect~\ref{sect:neatmmod} in the Appendix). This approach involved accounting for the available size and albedo constraints for Makemake itself, and also considering additional components. The models are: 1) a single terrain Makemake; 2) a single terrain Makemake with a dark moon; 3) a double terrain Makemake featuring a mix of bright and dark areas. The results are presented in Fig.~\ref{fig:neatm3}. We note that reflected light has a nearly negligible contribution to our thermal measurements. 

A single terrain model (Case~1, Fig.~\ref{fig:neatm3}a) can fit the long-wavelength part of the Spectral energy distribution (SED) very well, 
indicating that the bright terrain component on Makemake is well described by this model. However, at $\lambda$\,$\leq$\,100\,$\mu$m the model deviates from the observed values, differing by 1-2 orders of magnitude. 

Incorporating a dark satellite into the model (Case 2, Fig.~\ref{fig:neatm3}b) improves the fit at shorter wavelengths and provides an acceptable fit at 70\,$\mu$m. However, to fit the mid-infrared flux densities, a very dark ($p_V$\,$\lesssim$\,0.02) satellite with extreme thermal properties ($\eta$\,=\,0.34) has to be considered, as suggested by \citet[][labelled in the figure as L17]{Lellouch2017}. 
Even with this adjustment, the flux density measured by JWST/MIRI F1800W (18\,$\mu$m) is still significantly underestimated by an order of magnitude. If a second -- hypothetical -- extremely large and dark satellite is added to the system, the 18\,$\mu$m flux density can indeed be fitted ('giant moon', magenta curve in Fig.~\ref{fig:neatm3}b). However, this additional body should have a diameter of $\sim$1200\,km, comparable to the size of Makemake, and a dark and rough terrain (p$_V$\,=0.04, $\eta$\,=\,0.6), which is clearly not supported by any measurements. Furthermore, in this scenario, the flux densities at all longer wavelengths are significantly overestimated, a result that is corroborated by thermophysical model calculations (see below). 

Similarly, considering a mixture of bright and dark terrains (Case 3, Fig.~\ref{fig:neatm3}c) can improve the fits at shorter wavelengths, but cannot fit the mid-infrared and far-infrared observed flux densities at the same time, and the 18\,$\mu$m flux density clearly requires additional surfaces, as in the case of the extra giant dark body above. 

To confirm the validity of the previous NEATM calculations we also performed thermophysical model calculations \citep{Lagerros1996A&A...310.1011L,Lagerros1998A&A...332.1123L}, using a wide range of thermophysical model parameters (surface roughness, thermal inertia, spin properties), to obtain radiometric size and albedo which match the observed flux densities. 
This showed, in agreement with the NEATM results, that while for the longer wavelengths the occultation size is matched, a very large (D\,=\,4000-6000\,km equivalent size) and very dark (p$_V$\,$\leq$\,0.05) Makemake (or another body) is needed to obtain the observed mid-infrared flux densities, which is clearly incompatible with the observations. 


As a conclusion, thermal emission of solid surfaces of airless bodies, heated by the solar irradiation alone, cannot fully reproduce the observed infrared SED of Makemake. In particular, no model could fit the mid-infrared excess seen by JWST/MIRI in the F1800W band.  

A detailed analysis (Appendix, Sect.~\ref{sect:contamination}) also shows that neither reasonable contaminating sources (galaxy, bypassing main belt asteroid) nor photometric colour corrections or other instrument issues could feasibly explain the observed mid-infrared flux densities. Therefore, the source of the mid-infrared excess must be on or around Makemake. Below, we propose two possible scenarios to explain this very prominent mid-infrared excess. 

\section{Makemake with `hot spot' \label{sect:hotspot}}


Material from subsurface activity of icy bodies may reach the surface and cause excess temperatures. One prime example is Enceladus \citep{Spencer2006} where Cassini detected 3--7\,GW of thermal emission from the south polar troughs at temperatures up to at least 145\,K, from an equivalent area of $\sim$350\,km$^2$ ($\sim$10\,km radius). We may assume that the origin of Makemake's mid-infrared excess is similar. 

\begin{figure}[ht!]
    \centering
     \includegraphics[width=0.96\columnwidth]{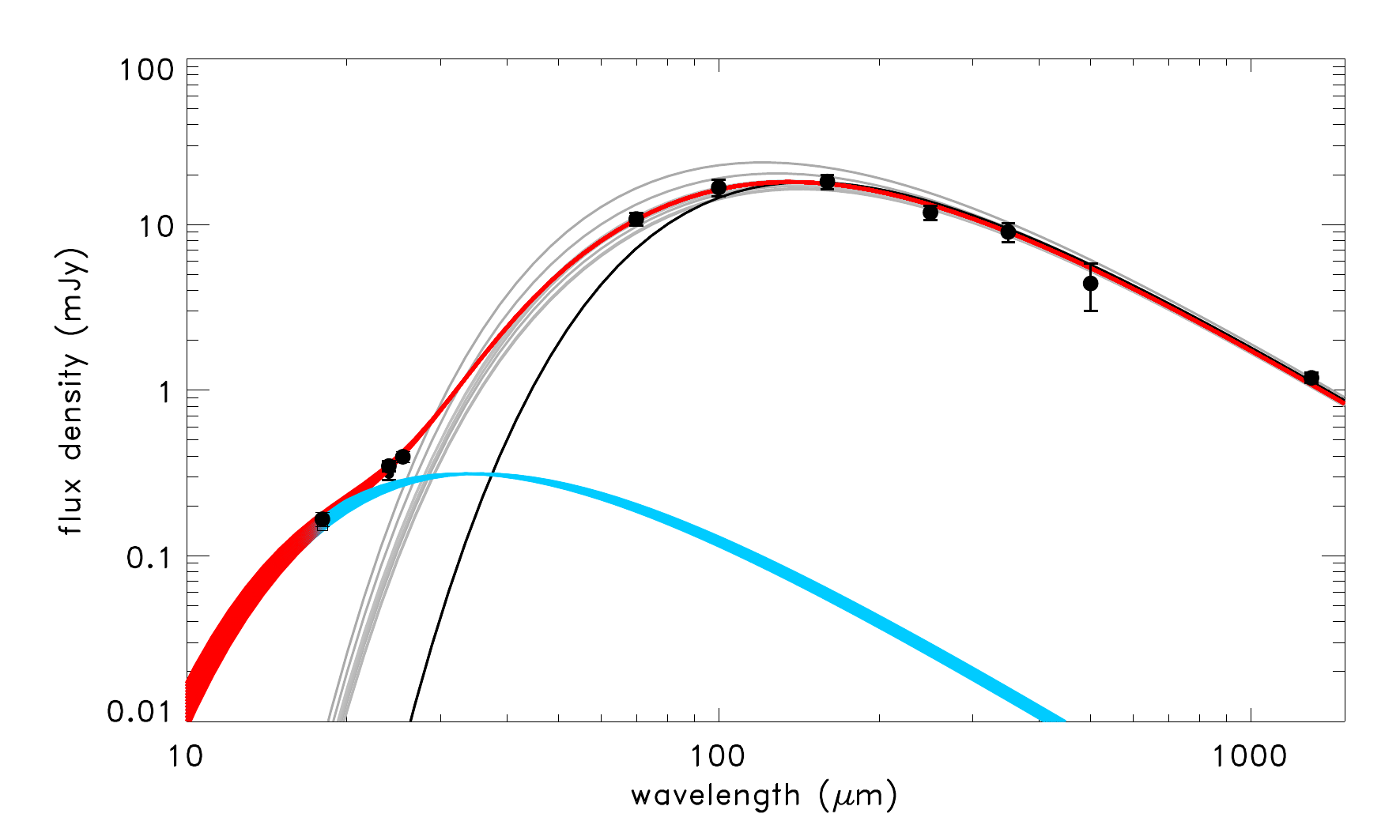}
    \caption{Spectral energy distribution assuming an additional `hot spot' on the surface. Gray curves correspond to a single terrain Makemake + a dark satellite with different albedos. The light blue curves correspond to black bodies in the T$_s$\,=\,147$\pm$5\,K range. The red curves are the sum of the blue curves and the `median' Makemake + satellite curve, assuming a satellite with p$_{V,S}$\,=\,0.04 and D$_s$\,=\,179\,km.}
    \label{fig:hotspot}
\end{figure}

In this scenario, we fitted the spectral energy distribution assuming a single-terrain Makemake and a `median' satellite contribution ($p_V$\,=\,0.04), as described in Sect.~\ref{sect:neatmmod}, and using a `hot spot' assuming that it has a spectral energy distribution of a single temperature black body (Fig.~\ref{fig:hotspot}).
For this additional component we obtain a best fitting black body temperature of T$_s$\,=\,147$\pm$5\,K, and the corresponding area has an equivalent radius of $r_s$\,=\,10.0$\pm$0.5\,km, i.e. $\sim$0.02\% of the apparent disk of Makemake. 
\citep[Wright Mons on Pluto, a suspected cryovolcano, has a caldera of $\sim$5\,km in diameter,][]{White2017,Singer2022}.
 As the Spitzer/MIPS 24\,$\mu$m flux densities show a small variation only ($\leq$20\%, see Sect.~\ref{sect:thermal}), in this scenario the hot spot should be continuously visible and related to a continuously visible (polar) region. In this case, depending on the actual pole orientation, the actual true area may be significantly larger and still have an equivalent radius of $r_s$\,=\,10\,km due to projection effects. In the case of a perfectly equator-on configuration a polar cap of $\sim$3.5\degr\, radius in latitude would be required to produce the area and excess power observed, covering $\sim$1\% of the total surface of Makemake. 
A 147\,K black body corresponds to a radiated power surface density of $\sim$26\,W\,m$^{-2}$, significantly larger than the typical $\lesssim$1\,W\,m$^{-2}$ on the other regions on Makemake's surface where the power output is determined by the solar irradiation. 
We can also calculate the total power radiated by this hot spot, assuming the two limiting cases. For an area with an equivalent radius of $r_s$\,=\,10.0$\pm$0.5\,km the corresponding total power is $P_{tot}$\,=\,8.3$\times$10$^9$\,W (coincidentally, quite similar to the values obtained for Enceladus), while it is 1.5$\times$10$^{11}$\,W for the 3.5\degr\, polar region assumed above. This can be compared with the 8$\times$10$^{11}$\,W total power received by Makemake from the solar irradiation. 
With our current (lack of) knowledge about Makemake's surface we cannot identify an obvious source that could be responsible for this high temperature region. 

 However, \citet{Glein2024} suggested that the D/H ratio of Makemake observed by JWST \citep{Grundy2024} is consistent with an abiotic and/or thermogenic origin of methane. These processes require hydrated rocky cores with high interior temperatures of 420-670\,K, which may have been reached early in Makemake's history, as 
supported by the available thermal evolution models. 
On such an internally evolved world, the origin of surface methane may be cryovolcanic outgassing from an interior water ocean, or solid state convection followed by clathrate degassing. Both processes can provide a substantial heat flow to the surface from the warm interior \citep{Grundy2024}. 
If the hot spot discussed here is the reason behind the mid-infrared excess emission, then Makemake is only the fourth known solid planetary body -- after Earth, Io and Enceladus -- that is sufficiently geologically active for its internal heat to be detected by remote sensing.

There are several scenarios which could potentially explain elevated surface temperatures. 
When trying to find some analogy with other active bodies with similar surface compositions, Triton has known surface activity in the form of plumes which are usually explained as eruptive processes \citep[see the summary by][]{Hofgartner2022}. Possible scenarios to explain this feature include versions of a solid state greenhouse effect model where a layer of nitrogen ice is more transparent to the incident solar radiation than the emitted thermal radiation, leading to a temperature increase within or at the bottom of the ice layer. While in this particular model \ce{N2} is assumed, other ices, like \ce{CH4} may play a similar role, also on Makemake. However, the solid state greenhouse effect can produce excess temperatures of $\lesssim$20\,K only \citep{Brown1990}, and the observed $\sim$150\,K associated with Makemake's mid-infrared excess is notably above the melting temperatures of N$_2$ and CH$_4$ ices (63\,K and 91\,K, respectively), essentially ruling out this scenario.


Among classical cryomagmas, a solution of \ce{H2O-NH3-CH3OH} has the lowest freezing temperature of $\sim$150\,K \citep[see e.g.][]{Kargel1998}, very similar to the excess temperature observed for Makemake. 
However, the observed temperature may rather be a lower limit on the actual temperatures on warm fissures or flows if they are not a continuous sheet, but are instead small-scale geologic features interspersed by cooler terrains. This effect is observed on Enceladus, where heat is localized along four "tiger stripes". If sources of anomalous thermal emission on Makemake exhibit similar behavior, then solutions with different solute concentrations, like water containing \ce{NH3} or dissolved salts, would also be possible.



In light of the evidence for a geophysically active interior, as proposed by \citet{Glein2024} to explain methane formation, and the need to maintain methane on the surface against escape and photochemical/radiolytic loss processes \citep{Grundy2024}, cryovolcanism may support the scenario involving a subsurface water ocean rather than one dominated by solid-state convection \citep[although, see][]{Nimmo2023}. 
Should the excess temperature be associated with cryovolcanism, observation of the variation of the excess temperature with time or with rotational phase may provide more insights into the actual process. 

Further insight into the plausibility of current geophysical activity can be elucidated by examining the energy balance. While the density of Makemake is currently uncertain \citep{Parker2018}, if we assume that it is composed of $\sim$70\% rock and $\sim$30\% water \citep[like most Kuiper belt objects;][]{Bierson2019}, then we can estimate rock and water masses of $\sim$2.2$\times$10$^{21}$\,kg and $\sim$0.9$\times$10$^{21}$\,kg, respectively. For a radiogenic heating rate of 5.6\,pW per kg of rock \citep{Desch2009}, we calculate $\sim$12\,GW of current heat production in the interior. This is sufficient to explain our lower limit for the excess thermal emission ($>$8\,GW). However, Makemake should also be releasing heat via thermal conduction spread over its surface. Thus, heat production is probably operating at a deficit, and Makemake would not be at steady state if it is releasing such a large quantity of heat. The source of excess heat may reside in a subsurface water ocean that is now freezing. As an example, let us consider what the rate of heat release would be if a mass of liquid water comprising ~1\% of the total water inventory has been freezing over the past 10$^6$-10$^7$\,yr. The amount of latent heat released is $\sim$3$\times$10$^{24}$\,J, which corresponds to $\sim$10-100\,GW. This seems broadly consistent with the amount of heat needed to explain the thermal anomaly that we identified. Future modeling will need to assess whether a subsurface ocean can persist until today, how liquids may be brought to the surface of Makemake, and why a burst of liquid water freezing happens to be taking place at the present time.

One important aspect of the cryovolcanic senario is that this kind of activity may result in a significant amount of material that could potentially produce a global atmosphere. This has not been identified in occultation measurements, with an upper limit of 4-12\,nbar surface pressure \citep{Ortiz2012}. At the current 52\,au heliocentric distance, however, volatiles that may form Makemake's atmosphere, N$_2$ and/or CH$_4$, quickly recondense after being released, and may just form a local atmosphere or plum around the active region \citep{Hofgartner2019}.

\section{Dust around Makemake \label{sect:ring}}

One may also assume that the origin of the mid-infrared excess is high-temperature dust made of small grains, as small grains tend to overheat due to their low emissivities, with the actual dust temperature depending on the composition and grain size \citep[see e.g.][]{Henning1996}. These small grains may reach much higher temperatures than the $\sim$40\,K equilibirium temperature at Makemake's heliocentric distance. 

We performed dust temperature calculations for the case of Makemake in Sect.~\ref{sect:dusttemperature}. 
As we have seen above, temperatures around 150\,K are required to explain the observed prominent mid-infrared excess. 
Dust temperatures for a specific grain composition reach the maximum around a grain size of $\sim$100\,nm, but temperatures around 150\,K are reached for graphite or carbon grains only, indicating that likely carbonaceous composition and small grain sizes are required to be able to explain the mid-infrared excess with diffuse dust. 

Recent studies found ring systems around Centaurs and trans-Neptunian objects including Chariklo, Haumea and Quaoar \citep{BragaRibas2014,Ortiz2017,Morgado2023,Pereira2023}, suggesting that these rings may be common around outer Solar system bodies, and here we assume that diffuse dust around Makemake may have a similar form.  

The chords of the April 2011 occultation \citep[see Fig.~2 in][]{Ortiz2012} run through Makemake's limb roughly in the east-west direction, and an ellipsoidal fit provides an apparent ellipse with a position angle (long axis vs. the north direction) of 9\degr$\pm$24\degr. As the orbit of Makemake's satellite is seen close to edge-on (A. Parker, priv. comm.), it is possible that there is a ring seen at a low opening angle (B\,$\lesssim$\,15\degr, as allowed by the occultation chords) and with a position angle similar to that of the fitted ellipsoidal limb. This ring could have avoided discovery in the occultation measurements. 


We performed radiative transfer model calculations using a simple ring model (Sect.~\ref{sect:ringrad} in the Appendix).
Our results show that due to their lower dust temperatures the SED of larger grains (s\,$\gtrsim$500\,nm) cannot fit the observed mid-infrared (18-25\,$\mu$m) emission of Makemake. This is also the case for some of the small (100 or 200\,nm) grains. If we use the respective SEDs normalized to the measured F1800W data, the SEDs of most materials overestimate the 24 and 25\,$\mu$m data (e.g. olivine, pyroxene, water ice). On the other hand, the SEDs of 100-200\,nm graphite grains considerably underestimate the observed 24 and 25\,$\mu$m flux densities (i.e. these grains are `too hot'). 
Among the materials we investigate here carbonaceous grains \citep{Zubko1996} with grain sizes 100-200\,nm, or graphite grains with grain sizes between 200 and 500\,nm can fit all mid-infrared data simultaneously  (without violating the JWST/MIRI F560W detection of the reflected light), considering some possible contribution from other components (dark moon, dark terrain) in the system in addition to the cold and bright surface of Makemake.

\begin{figure}[ht!]
    \centering
    \includegraphics[width=\columnwidth]{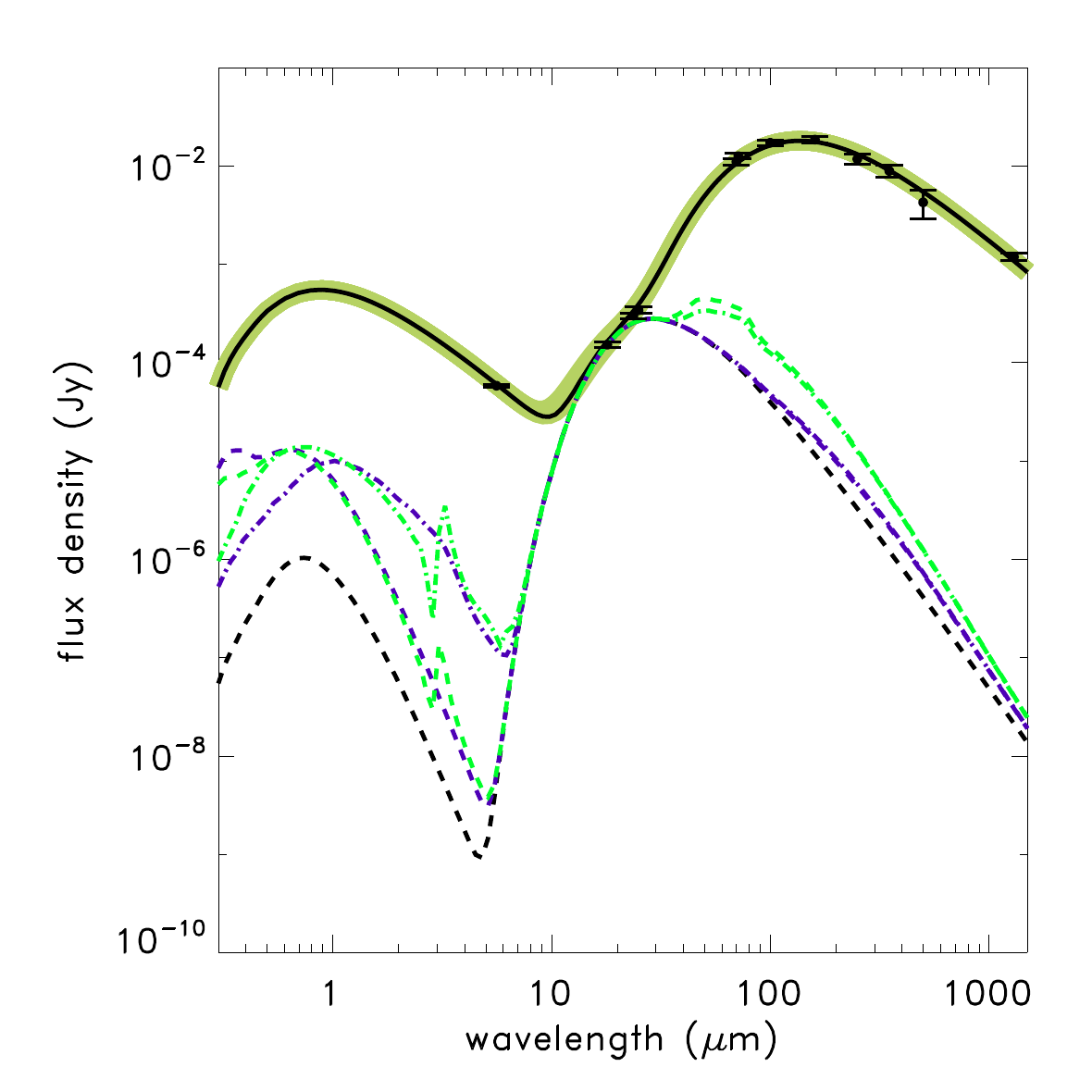}
    \caption{Spectral energy distribution of Makemake's reflected light and thermal emission. Black symbols represent the mean measured flux densities. 
    The green stripe shows the SED of the `hot spot' models, compatible with the observations, as described in Sect.~\ref{sect:hotspot} (note that the stripe was made somewhat wider than it is in reality, for a better visibility). It was calculated assuming that Makemake has a double-terrain with `Quaoar-like' dark terrains, and reflected light with a mid-infrared albedo of $p_{5.6}$\,=\,0.8. The black dashed curve is the SED of 100\,nm carbonaceous grains that together with Makemake's contribution (solid black curve inside the green stripe) matches the observed flux densities very well.
    The coloured dashed/dash-dotted curves correspond to the emission of dust when other types of dust grains are also included, in addition to the same amount of 100\,nm carbonaceous dust as in the pure case. The additional material has a mass of 100$\times$ the mass of 100\,nm carbonaceous grains in all cases. 
    The additional components are: 
    Purple dashed curve: pyroxene with 100\% Mg-content, 100\,nm grain size; 
    Purple dash-dotted curve: pyroxene with 100\% Mg-content, 1\,$\mu$m grain size; 
    Green dashed curve: amorpous water ice, 100\,nm grain size; 
    Green dashed-dotted curve: amorpous water ice, 1\,$\mu$m grain size}. 
    \label{fig:dustmixing}
\end{figure}

In Fig.~\ref{fig:dustmixing} we present the spectral energy distribution of Makemake with a ring made of 100\,nm-sized carbonaceous grains (black solid curve), with an additional contribution of a double-terrain model using Quaoar-like secondary terrain (see Sect.~\ref{sect:neatmmod}), {however, using a dark satellite model instead of double terrain provides very similar results.}
This particular model gives a very good match to the observed flux densities, and is practically indistinguishable from the best fit `hot spot' model (Sect.~\ref{sect:hotspot}, and green stripe in Fig.~\ref{fig:dustmixing}). If the ring was made exclusively of these very small grains, the optical depth of the ring in the visible range would be $\tau$\,$\approx$\,0.1, assuming a thin and narrow disk, and a ring width of 10\,km and ring radius of r\,=\,4300\,km, corresponding to the 3:1 spin-orbit resonance, similarly to those found around other small bodies \citep{BragaRibas2014,Ortiz2017,Morgado2023}. Note that the optical depth depends on the actual ring width chosen, and smaller for a wider ring. 

{ All known dusty rings of the giant planets are associated with denser rings or small moons, which also serve as dust sources via micrometeorite impacts, resupplying the small grains in the ring, which are quickly lost from the system due to non-gravitational effects \citep{Hedman2018}.}
If Makemake's dusty ring exists, it is likely that it contains material with a wider range of grain sizes, and maybe a mixture of compositions.  
Using a few simple examples we tested how the presence of other materials would modify the SED of the ring. Due to the predominantly cold temperatures of Makemake's surface ($\sim$40\,K) larger grains, which have similarly low dust temperatures, do not contribute notably to the mid-infrared emission, and they have to `compete' with Makemake's cold thermal emission to be visible in the far-infrared. This is also true for the large ($\gtrsim$100\,$\mu$m) particles that make up the classical giant planet rings.
In Fig.~\ref{fig:dustmixing}, we present a few examples in which we added the contribution of different grains to that of 100\,nm carbonaceous grains. The contribution of the additional material is 100-times the mass of the 100\,nm carbonaceous grains in all cases. In all these cases the mid-infrared emission remains dominated by the 100\,nm carbonaceous grains, with a negligible contribution from the other materials. 
While these additional materials dominate the reflected light and in some cases the far-infrared emission of the ring, they do not change the overall SED due to the dominance of the reflected light and thermal emission from Makemake itself. One important aspect is, however, that adding these materials pushes the visible range opacities to $p$\,$\lesssim$\,1, which makes the ring easily detectable by stellar occultations, when the chords are cutting through these structures. 

Small grains will be strongly affected by solar radiation through radiation pressure and Poynting-Robertson effects \citep{Burns1979}. We performed a detailed calculation of the particles lifetime (see Sect.~\ref{sect:grainremoval}), using a dynamical model considering solar radiation pressure effects and assuming that the grains start orbiting Makemake in circular orbits. 
We estimate that the lifetime of the smallest carbonaceus grains -- which are suspected to be responsible for the mid-infrared excess emission -- is $\sim$10\,yr, depending on the starting semi-major axis (see Fig.~\ref{fig:lifetime}).
We note, however, that a small shepherding moonlet may help to stabilize the orbits of the ring particles and significantly extend the lifetime of the grains, like it has been proposed in the case of Chariklo \citep{Sickafoose2024,Salo2024}. 
All other timescales, including the collisional timescale of dust grains and the decay due to the Poynting-Robertson drag \citep{Burns1979,MD2000} are orders of magnitude longer than the radiation pressure timescale for small grains, and likely do not play an important role here. 


The ring proposed to explain Makemake's mid-infrared excess here would be a new type of ring in the Centaur and trans-Neptunian regions. The thermal emission of the Haumea, Chariklo or Quaoar rings does not show a similar, strong mid-infrared excess emission \citep{Muller2019,Lellouch2017,Kiss2024}, however, recent results from stellar occultations suggest that very small grains may dominate some small body rings in the outer Solar System (Santos-Sanz et al. 2024, submitted).
While the `classical' rings of Saturn and Uranus are known to be mostly mm-to-cm sized grains or pebbles \citep{2018prs..book...51C,2018prs..book...93N}, the thermal emission of Phoebe ring around Saturn \citep{Verbiscer2009} is dominated by small grains, and it is characterised by a very steep size distribution law, as obtained by WISE and Spitzer measurements \citep[size distribution power law index of q\,=\,4--6,][]{Hamilton2015}. In the case of Phoebe ring dust particles are originated from the satellite Phoebe itself from micrometeorite, or larger impacts, and are thought to be responsible for the dark material on the leading hemisphere of Iapetus \citep{Tamayo2011}. Similarly, an additional, small inner satellite may be responsible for the ring material in the case of Makemake. As the lifetime of the small grains in the putative Makemake ring is on the order of a decade, {in the case of a  single event a fading should have been observed in the last $\sim$20\,yr, covered by mid-IR observations. However, the earlier Spitzer/MIPS and the recent JWST/MIRI data show compatible flux densities, indicating that} there may be a continuous replenishment of dust that keeps the ring material continually observable. Concerning the composition of dust grains, carbon is ubiquitous in the outer Solar system. Submicron cometary dust is dominated by amorphous carbon \citep{Harker2023} and, as we have shown above, due to their unique optical properties very small carbonaceous dust grains may be the dominant source of the mid-infrared thermal emission even in the presence of other types of grains. 
{Using our radiative transfer calculations we estimate that assuming solely 100\,nm carbonaceous grains the total mass of the ring is $\sim$3$\times$10$^{6}$\,kg, equivalent to the mass of a body with $\sim$10\,m radius. This is likely a lower limit as larger grains may contribute significantly without notably modifying the spectral energy distribution of the system, as it was shown above. Assuming the presence of larger grains with hundred times the mass of small grains, the mass of the ring could be 3$\times$10$^{6}$\,kg\,$\lesssim$\,$M_r$\,$\lesssim$3$\times$10$^{8}$\,kg.
With a short lifetime of $\sim$10\,yr of the very small grains, a rate of $\dot{M_r}$\,$\gtrsim$3$\times$10$^5$\,kg\,yr$^{-1}$ is required for replenishment, and a probably higher amount considering the whole scale of particle sizes which may originate from small moons or collisions between large ring particles. 
}

\section{Conclusions \label{sect:conclusions}}

We have shown that the trans-Neptunian dwarf planet (136472)~Makemake exhibits a prominent mid-infrared excess that cannot 
be explained by the thermal emission of solid bodies irradiated only by the Sun, at the heliocentric distance of Makemake.

We proposed two separete scenarios to explain this mid-infrared excess: a hot spot powered by cryovolcanism, or a ring made of very small carbonaceous grains. 
{Interestingly, these two phenomena may be interconnected. The material of Saturn's E ring originates from Enceladus' water geysers \citep[see e.g.][for a summary]{Hedman2018}, and similar processes may supply material to a ring around Makemake. In addition, Saturn's E ring is also dominated by sub-micron sized grains. If these processes can put small carbonaceous or graphite grains around  Makemake (e.g. in addition to water ice grains) the actual observed infrared excess may be a result of a combination of the two phenomenon. 

The Spitzer/MIPS partial light curve (Sect.~\ref{sect:thermal}) indicated a small amplitude (16\%) variation at 24\,$\mu$m, however, a flat light curve could not be excluded. 
}
Additional mid-infrared (10-25\,$\mu$m) measurements sampling Makemake's thermal emission at multiple subobserver longitudes may confirm this rotational variation, which could be a strong indication that the excess emission at least partly comes from Makemake's surface. However, a rotationally constant excess emission does not automatically prove the existence of a ring. Additional mid-infrared observations may also show whether the excess -- both the intensity and the associated temperature -- has changed since the latest JWST/MIRI measurements in January 2023. This is expected in the ring scenario if the responsible small dust grains were created in a single event. However, in the case of a hot spot, the excess may also change due to changes in the underlying processes (e.g. cooling cryolava). 
Future occultation measurements may indeed help to solve the ring vs. hot spot puzzle. 

\clearpage

\section*{Acknowledgements}

The research leading to these results has received funding from the K-138962 and TKP2021-NKTA-64 grants of the National Research, Development and Innovation Office (NKFIH, Hungary). P.S-S. acknowledges financial support from the Spanish I+D+i project PID2022-139555NB-I00 (TNO-JWST) funded by MCIN/AEI/10.13039/501100011033. P.S-S. and J.L.O. acknowledge financial support from the Severo Ochoa grant CEX2021-001131-S funded by MCIN/AEI/10.13039/501100011033. C.R.G. was supported by the NASA Astrobiology Institute through its JPL-led team entitled Habitability of Hydrocarbon Worlds: Titan and Beyond.
This work is based in part on observations made with the NASA/ESA/CSA James Webb Space Telescope. The data were obtained from the Mikulski Archive for Space Telescopes at the Space Telescope Science Institute, which is operated by the Association of Universities for Research in Astronomy, Inc., under NASA contract NAS 5-03127 for JWST. We thank our reviewer for the comments and corrections. 

\vspace{5mm}
\facilities{TESS \citep{Ricker2015}, JWST/MIRI \citep{Wright2023}, Spitzer/MIPS \citep{MIPS}, Herschel/PACS \citep{PACS}}


\software{FITSH \citep{Pal2012}, OPTOOL \citep{optool}, RADMC-3D \citep{radmc}}


\newpage



\setlength{\bibsep}{0pt}

\bibliography{bib}
\bibliographystyle{aasjournal}




%


\newpage

\appendix
\twocolumngrid


\section{Visible range light curves \label{sect:period}}

\paragraph{TESS data \label{sect:tess}}

\begin{figure}[ht!]
    \centering
    \includegraphics[width=\columnwidth]{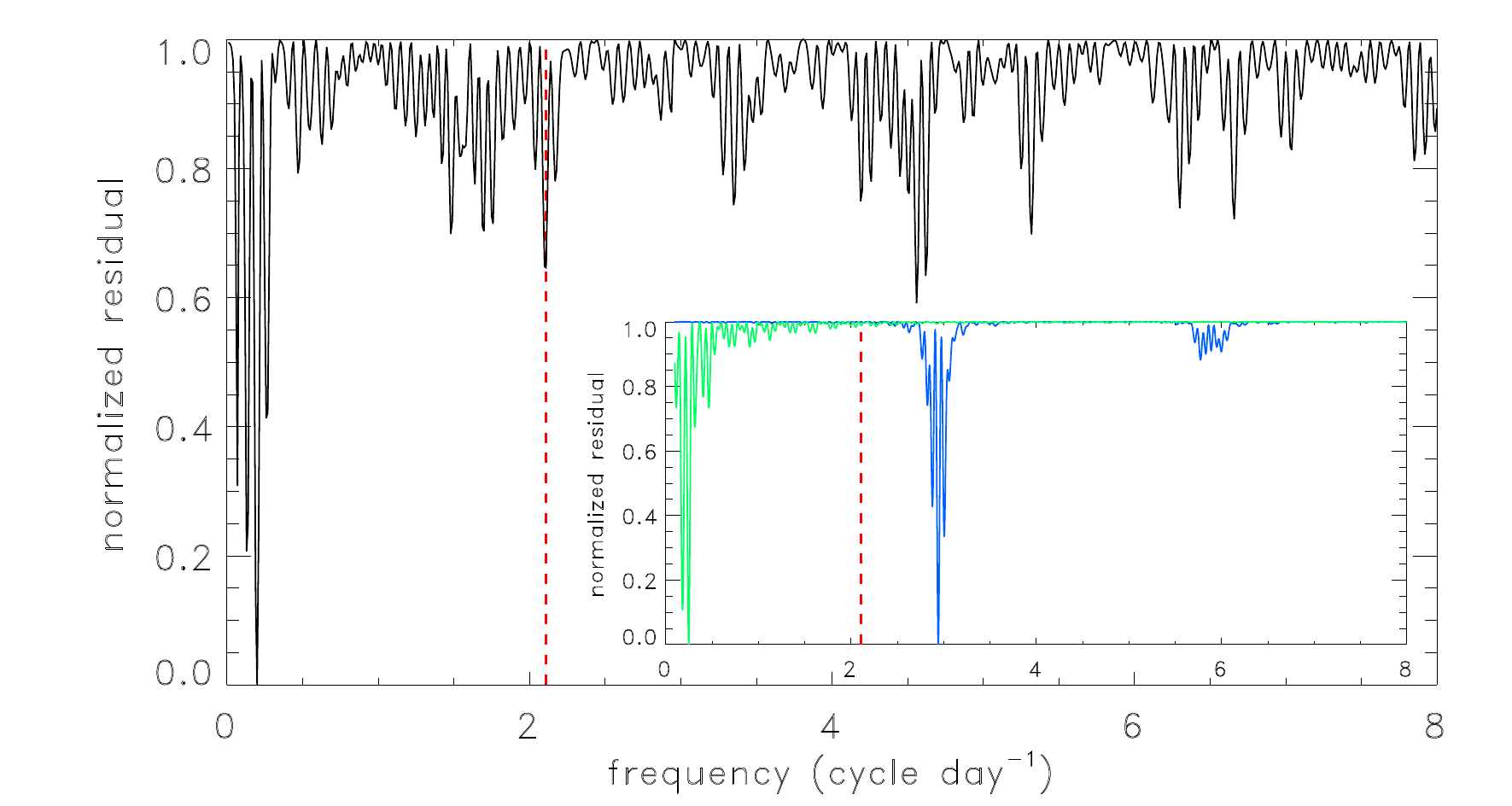}
    \includegraphics[width=\columnwidth]{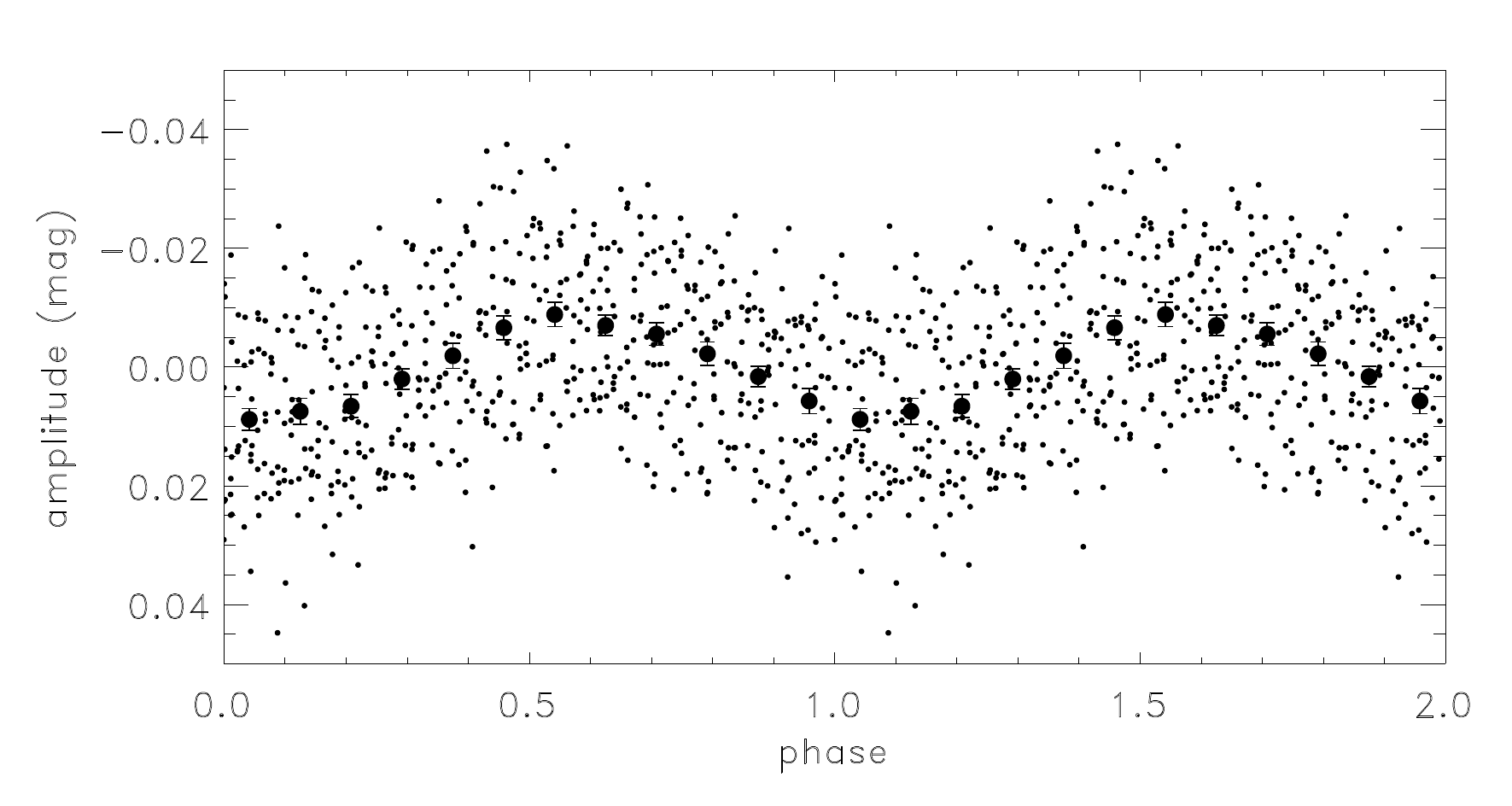}
    \includegraphics[width=\columnwidth]{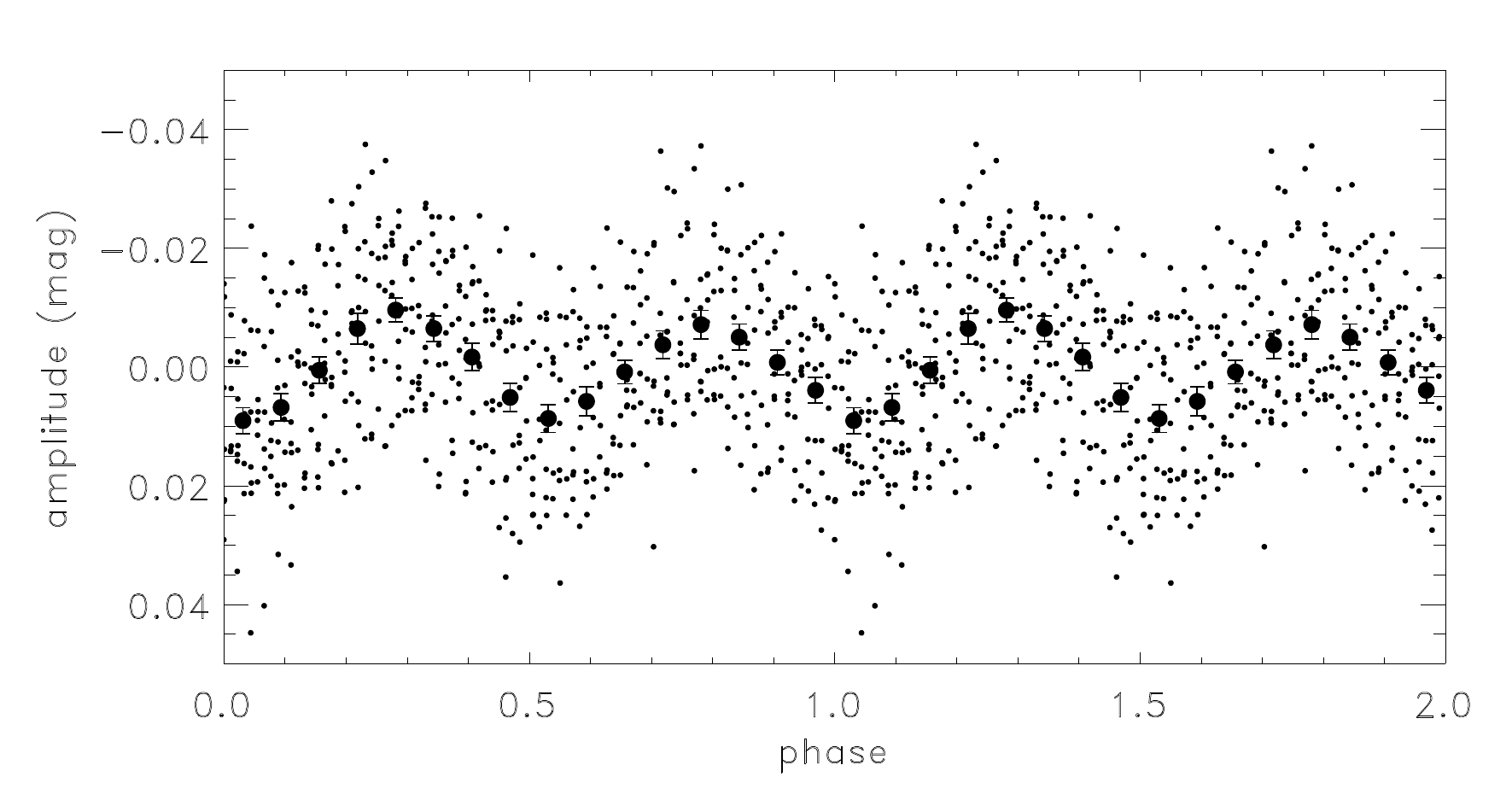}
    \caption{Top: Phase dispersion vs frequency of the Makemake TESS data. The insert shows the residual spectrum derived using the position of aperture centres with respect to the actual pixel's edge, in the 'X' (blue) and 'Y' (green) directions. 
    In both the main figure and the insert, the vertical, dashed red line is at the frequency of f\,=\,2.105\,c/d (P\,=\,11.401\,h), associated with the prominent peak in the residual spectrum, very close to the 11.41\,h period identified by \citet{Hromakina2019}. 
    Middle: Light curve folded with the period of P\,=\,11.401\,h. Black dots represent the binned data; the red curve is the best-fit sinusoidal.  
    Bottom: Light curve folded with the double, P\,=\,22.802\,h period. 
    }
    \label{fig:tesslc}
\end{figure}

The Transiting Exoplanet Survey Satellite \citep[TESS,][]{Ricker2015} observed Makemake in Sector 23, between March 19 and April 15, 2020, and in Sector 50 between March 26 and April 22, 2022. The Sector 50 data is heavily affected by straylight, therefore we used the Sector 23 measurements only. Basic data reduction was performed with the TESS reduction pipeline developed for solar systems targets, as described in \citet{Pal2020}. Due to the large pixel size of 21\arcsec\, and the slow motion of Makemake in the sky, the photometry is affected by the relative position of the measuring aperture and the edge of the actual pixel, introducing characteristic frequencies in the residual spectrum, related to the X and Y direction of motion of the target through the field of view. To account for this effect, we used the measured brightness values of Makemake versus the X and Y pixel fractions, and produced phase dispersion spectra in a way similar to that of the `normal' lightcurve (brightness versus time, see Fig.~\ref{fig:tesslc}). The X and Y residual spectra show prominent frequencies at f\,$\approx$\,3\,c/d and f\,$\leq$\,0.3\,c/d respectively,  {corresponding to the apparent motion and pixel-crossing times of Makemake in the specific directions}. One of the prominent peaks in the residual spectrum apart from the X- and Y-motion frequencies is at f\,=\,2.105$\pm$0.014\,c/d (P\,=\,11.401$\pm$0.076\,h), very close to the single peak solution obtained by \citet{Hromakina2019}. After correcting for the effect of the X- and Y-motion contribution we obtained the light curve folded by this f\,=\,2.105\,c/d frequency as presented in Fig.~\ref{fig:tesslc}, showing a peak-to-peak amplitude of $\Delta$m\,=\,0.018\,$\pm$0.002\,mag, notably smaller than the previous $\Delta$m\,=\,0.032$\pm$0.005\,mag amplitude obtained by \citet{Hromakina2019}. We also investigated whether a single or double peaked light curve is preferred, in the latter case using the double period, P\,=\,22.8\,h. Comparing the first and second half periods of the double peaked folded light curve using the method in \citep{Pal2016}, and also applying a Student's t-test \citep[see][]{Hromakina2019}, both methods agree that the two half periods are different at the $\sim$1.7\,$\sigma$ level, i.e. we cannot unambiguously distinguish between single and double peaked light curves. 

\paragraph{Gaia data \label{sect:gaia}}

Gaia data of Makemake is available in the third Gaia Data Release \citep{dr3}, accessible in the Gaia Science Archive\footnote{\url{https://gea.esac.esa.int/archive/}} through the \mbox{{\it gaiadr3.sso\_observation}} table. The table contains data obtained during the transit of the source on a single CCD, during a single transit. 
{For Makemake, the typical photometric errors of the individual data points are $\sim$0.4\% of the flux density}. 
More details about the SSOs in the Gaia DR3 are discussed in \citet{Tanga2023}. Gaia G-band data of Makemake was corrected for heliocentric and observer distance and phase angle, using spacecraft-centric data obtained from the NASA Horizons system \citep{1996DPS....28.2504G}. We applied a linear phase angle correction using the heliocentric and observer distance-corrected brightness values, and we used these reduced magnitudes for the period search. We applied both a residual minimization method \citep[see e.g.][]{Pal2015} and a Lomb-Scargle periodogram algorithm which resulted in essentially the same results. 
\begin{figure}[ht!]
    \centering
    \includegraphics[width=\columnwidth]{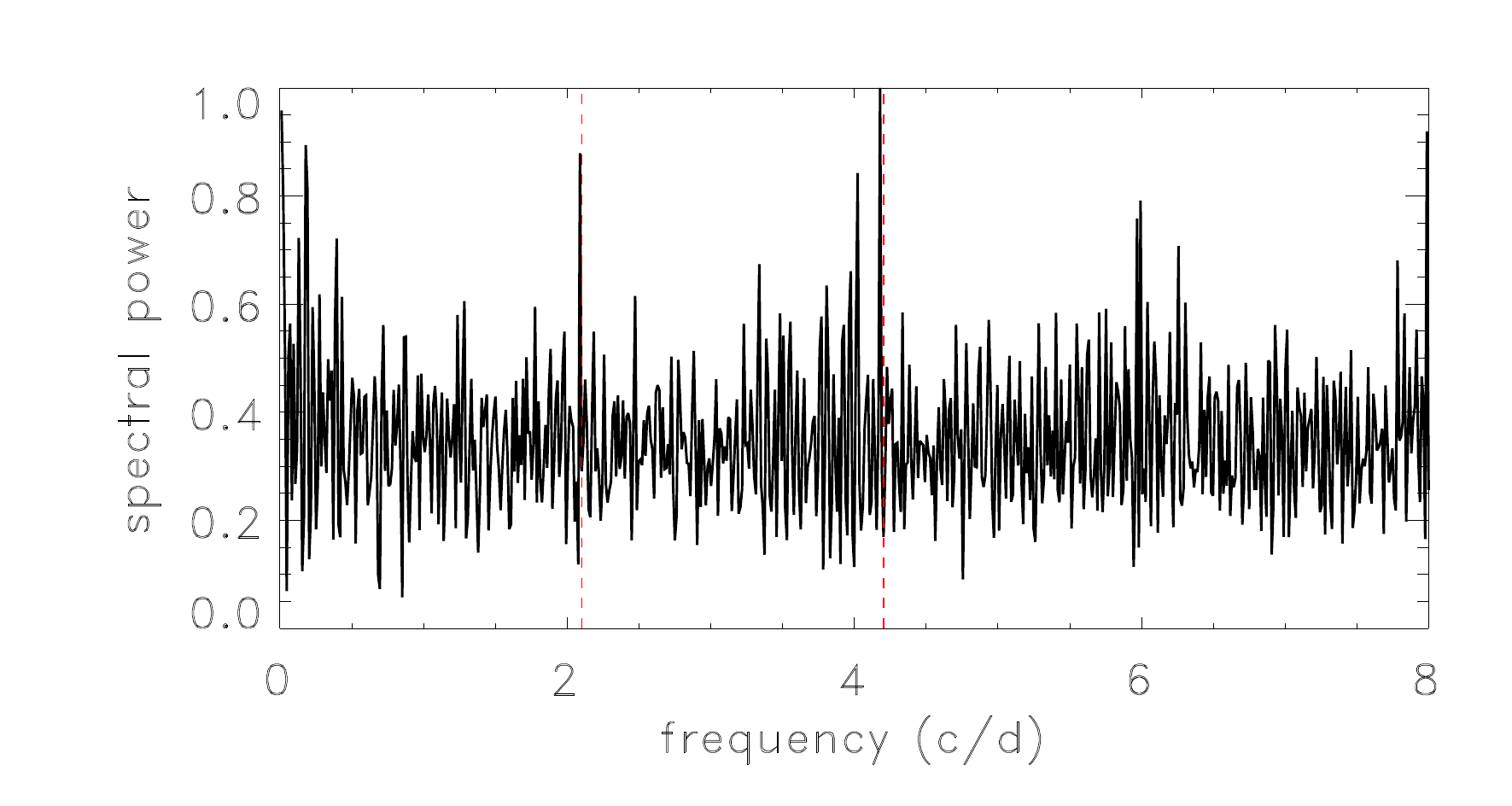}
    \includegraphics[width=\columnwidth]{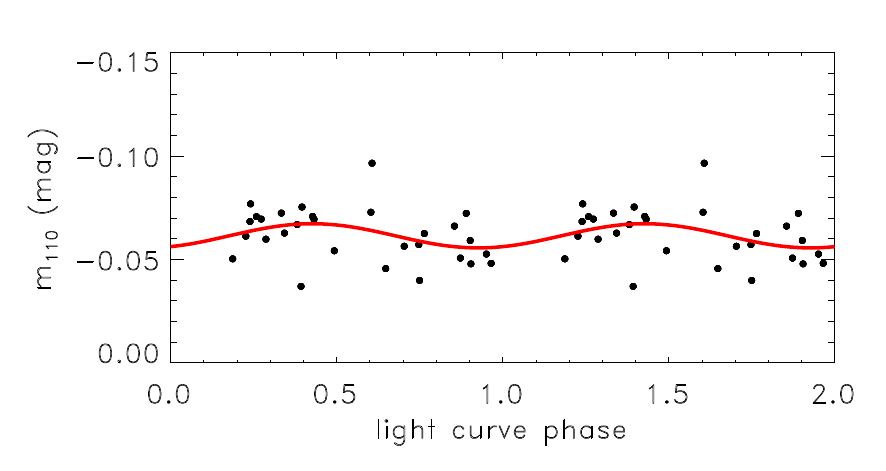}
    \caption{Top: Lomb-Scargle periodogram of the Makemake Gaia data. The vertical dashed lines mark the frequency corresponding to the 11.41\,h period, identified by \citep{Hromakina2019}, and the first overtone at 5.7\,h. 
    Bottom: Heliocentric and observer distance and phase angle corrected Gaia light curve, folded with the P\,=\,11.4767\,h period.}
    \label{fig:gaia}
\end{figure}

The Lomb-Scarge periodogram is presented in Fig.~\ref{fig:gaia}, and shows two primary peaks, one at f\,=\,2.0912$\pm$0.0015\,c/d or P\,=\,11.4767$\pm$0.0082\,h, and another peak at its first overtone at f\,=\,4.1793$\pm$0.0033\,c/d or P\,=\,5.7425$\pm$0.0045\,h with a peak-to-peak amplitude of $\Delta m$\,=\,0.012$\pm$0.002\,mag. 
{Both peaks have S/N\,$\approx$\,3 over the general spectral power fluctuation level, around the acceptable level of significance}. 
The P\,=\,11.4767\,h period is very close to the 11.41\,h single peak period identified by \cite{Hromakina2019} and also by us (11.40\,h) in the TESS data. The $\Delta m$\,=\,0.012$\pm$0.02\,mag Gaia amplitude is lower than the $\Delta m$\,=\,0.018$\pm$0.002\,mag obtained from the TESS data, and significantly lower than the $\Delta m$\,=\,0.032\,mag reported by \citet{Hromakina2019}. 

\medskip 

Some difference between the light curve amplitudes may be explained by the different filter transmission curves, i.e. the Gaia G-band filter is very wide and covers the spectral range between $\sim$400 and 900\,nm, while TESS sensitivity is mostly confined to the 600-1000\,nm range. Also, as Makemake orbits the Sun, the aspect angle of the spin axis changes and therefore the light curve amplitude changes as well. The \citet{Hromakina2019} study mixes observations with different instruments/filters from a wider period, including measurements from 2009 to 2017. This is a long time span, and consequently observations may include light curves with different amplitudes at different times -- this may at least partially explain the different half periods in the case of a double peak light curve, leading to the double peak light curve to be their preferred solution. 
Due to the low number of data points and the sporadic nature of the data, Gaia  measurements could not be used to distinguish between single and double  peaked light curves. 

\section{Infrared observations and data reduction \label{sect:thermal}}

\begin{table*}[ht!]
    \centering\small
    \begin{tabular}{ccccc|ccccc|c}
\hline
OBSID /      & Instrument &   JD$_{start}$    & JD$_{end}$ & t$_{tot}$ & r$_h$ & $\Delta$ & $\alpha$ & $\lambda$ & $\beta$ & bands\\
AORKEY & & & & (h) & (AU) & (AU) & (deg) & (deg) & (deg) & ($\mu$m) \\
\hline
\hline
\multicolumn{11}{c}{Herschel} \\
\hline 
1342187319	&	SPIRE	& 2455165.45072	& 2455165.46387	& 0.316	& 52.138 &	52.409 & 1.051 	& 176.544  	& 28.705	& 250, 350, 500	\\
1342187320	&	SPIRE	& 2455165.46453	& 2455165.47767	& 0.316	& 52.138 &	52.409 & 1.051 	& 176.544  	& 28.705	& 250, 350, 500	\\
1342187366	&	PACS	& 2455166.28238	& 2455166.30072	& 0.440	& 52.138 &	52.397 & 1.055 	& 176.552  	& 28.712	& 70, 160	\\
1342187367	&	PACS	& 2455166.30156	& 2455166.31990	& 0.440	& 52.138 &	52.397 & 1.055 	& 176.552  	& 28.712	& 100, 160	\\
1342187524	&	SPIRE	& 2455167.26799	& 2455167.28113	& 0.316	& 52.138 &	52.383 & 1.059 	& 176.561  	& 28.720	& 250, 350, 500	\\
1342187525	&	SPIRE	& 2455167.28179	& 2455167.29494	& 0.316	& 52.138 &	52.383 & 1.059 	& 176.561  	& 28.720	& 250, 350, 500	\\
1342197657	&	PACS	& 2455350.40924	& 2455350.41907	& 0.236	& 52.165 &	51.959 & 1.104 	& 174.615  	& 28.976	& 70, 160	\\
1342197658	&	PACS	& 2455350.41980	& 2455350.42964	& 0.236	& 52.165 &	51.959 & 1.104 	& 174.615  	& 28.975	& 70, 160	\\
1342197659	&	PACS	& 2455350.43037	& 2455350.44021	& 0.236	& 52.165 &	51.959 & 1.104 	& 174.615  	& 28.975	& 100, 160	\\
1342197660	&	PACS	& 2455350.44094	& 2455350.45078	& 0.236	& 52.165 &	51.959 & 1.104 	& 174.615  	& 28.975	& 100, 160	\\
1342197695	&	PACS	& 2455351.22786	& 2455351.23770	& 0.236	& 52.165 &	51.971 & 1.107 	& 174.613  	& 28.968	& 70, 160	\\
1342197696	&	PACS	& 2455351.23843	& 2455351.24826	& 0.236	& 52.165 &	51.971 & 1.107 	& 174.613  	& 28.968	& 70, 160	\\
1342197697	&	PACS	& 2455351.24899	& 2455351.25883	& 0.236	& 52.165 &	51.971 & 1.107 	& 174.613  	& 28.968	& 100, 160	\\
1342197698	&	PACS	& 2455351.25956	& 2455351.26940	& 0.236	& 52.165 &	51.971 & 1.107 	& 174.613  	& 28.968	& 100, 160	\\
1342198251	&	SPIRE	& 2455360.70866	& 2455360.73309	& 0.586	& 52.166 &	52.111 & 1.126 	& 174.612  	& 28.882	& 250, 350, 500	\\
1342198451	&	SPIRE	& 2455358.09508	& 2455358.11951	& 0.586	& 52.166 &	52.072 & 1.123 	& 174.609  	& 28.906	& 250, 350, 500	\\
\hline
\hline 
\multicolumn{11}{c}{Spitzer} \\
\hline
13803776  & MIPS &   2453542.19492  &  2453542.21199  & 0.410 &  51.883 & 51.869 & 1.122 & 169.577 & 29.026 & 24, 71.42 \\
13803264  & MIPS &   2453543.42211  &  2453543.43920  & 0.410 &  51.884 & 51.887 & 1.122 & 169.580 & 29.014 & 24, 71.42 \\
19179264  & MIPS &   2454256.98375  &  2454257.01009  & 0.632 &  52.001 & 51.523 & 0.984 & 171.788 & 29.281 & 24, 71.42 \\
19179008  & MIPS &   2454257.01442  &  2454257.04076  & 0.632 &  52.001 & 51.523 & 0.984 & 171.787 & 29.281 & 24, 71.42 \\
19178752  & MIPS &   2454257.04507  &  2454257.07141  & 0.632 &  52.001 & 51.524 & 0.984 & 171.787 & 29.281 & 24, 71.42 \\
19178496  & MIPS &   2454257.07573  &  2454257.10207  & 0.632 &  52.001 & 51.524 & 0.985 & 171.787 & 29.280 & 24, 71.42 \\
19178240  & MIPS &   2454257.11113  &  2454257.13748  & 0.632 &  52.001 & 51.525 & 0.985 & 171.786 & 29.280 & 24, 71.42 \\
19177984  & MIPS &   2454257.14178  &  2454257.16813  & 0.632 &  52.001 & 51.525 & 0.985 & 171.786 & 29.280 & 24, 71.42 \\
19177728  & MIPS &   2454257.17244  &  2454257.19878  & 0.632 &  52.001 & 51.525 & 0.985 & 171.786 & 29.280 & 24, 71.42 \\
19177472  & MIPS &   2454257.20309  &  2454257.22946  & 0.633 &  52.001 & 51.526 & 0.985 & 171.785 & 29.279 & 24, 71.42 \\
19177216  & MIPS &   2454257.24230  &  2454257.26867  & 0.633 &  52.001 & 51.526 & 0.986 & 171.785 & 29.279 & 24, 71.42 \\
19176960  & MIPS &   2454257.27676  &  2454257.30310  & 0.632 &  52.001 & 51.527 & 0.986 & 171.785 & 29.279 & 24, 71.42 \\
19176704  & MIPS &   2454257.30743  &  2454257.33377  & 0.632 &  52.001 & 51.527 & 0.986 & 171.785 & 29.279 & 24, 71.42 \\
19176448  & MIPS &   2454257.33807  &  2454257.36441  & 0.632 &  52.001 & 51.527 & 0.987 & 171.784 & 29.278 & 24, 71.42 \\
\hline
\hline
\multicolumn{11}{c}{ALMA} \\
\hline
        & ALMA  &   2457449.78750   & 2457449.80278 & 0.311 & 52.439   & 51.622 & 0.620 & 182.249 & 28.945 & 1300 \\
\hline
    \end{tabular}
    \caption{Summary table of thermal emission observations of Makemake, including Herschel/PACS and SPIRE, Spitzer/MIPS and ALMA observations.  The columns list the observation ID (AORKEY/OBSID), instrument, the start and end times in Julian date, target's heliocentric distance, observer to target distance, phase angle, and the central wavelengths of the filters/bands used. }
    \label{table:allmeas}
\end{table*}

\paragraph{MIPS 24/70\,$\mu$m measurements and 24\,$\mu$m partial light curve}

Makemake was observed with the MIPS photometer of the Spitzer Space Telescope \citep{MIPS} at two epochs, on June 20/21, 2005 and on June 5, 2007 (see also Table~\ref{table:allmeas}). The first epoch measurement consisted of two AORKEYs covering 0.82\,h while the second epoch covered 7.6\,h with 12 AORKEYs, in both cases using both the 24 and 70\,$\mu$m cameras. 
In the case of the 24\,$\mu$m maps we used the calibrated, post-BCD MAIC maps. 
The block of measurements made during the second epoch consists of 12 AOR keys, and altogether lasted $\sim$7.6\,h, covering a significant fraction of a full rotation, assuming the 11.4\,h period discussed above. At 24\,$\mu$m the source was clearly identified and bright with respect to the background in all individual AOR images. The sequence of images clearly show Makemake moving through the field, and indicate faint sources in the background that may notably affect the photometry. 
To account for this effect, we performed a background subtraction by dividing the 12 frames into two six-frame groups. We produced two 'shadow' images, the first one (S1) being the average image of the first two frames, and the second one (S2) being the average image of the last two frames. We used S2 to correct for the background of the first 6 frames, and S1 to correct for the background of the last 6 frames. Still, as Makemake moved $\sim$14\arcsec\, during the whole measurement, the distance between the closest frame and the corresponding shadow images are just $\sim$8\arcsec, only slightly larger than the $\sim$6\arcsec\, FWHM of the MIPS 24\,$\mu$m point spread function (PSF). 

\begin{figure}[ht!]
    \centering
    \includegraphics[width=0.95\columnwidth]{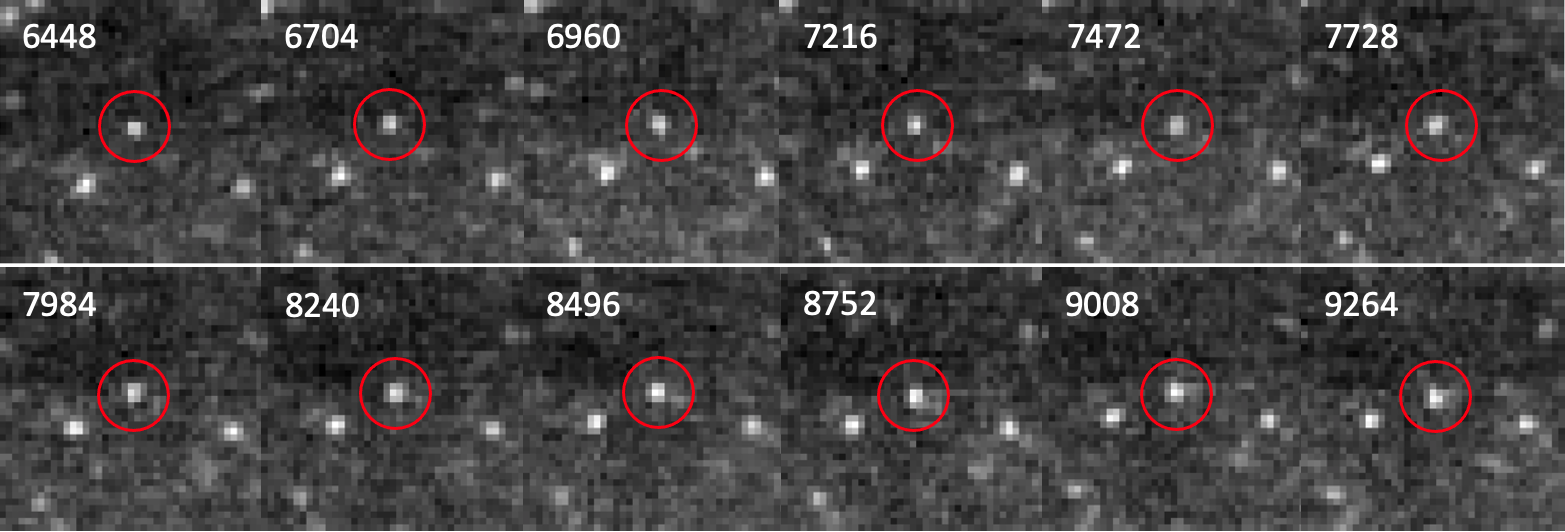}
    \caption{Spitzer/MIPS 24\,$\mu$m images of Makemake, with AORKEYs 19176448-19179264 (see Table~\ref{table:allmeas} for the flux densities derived).}
    \label{fig:mips24}
\end{figure}
\begin{figure}[ht!]
    \centering
    \includegraphics[width=\columnwidth]{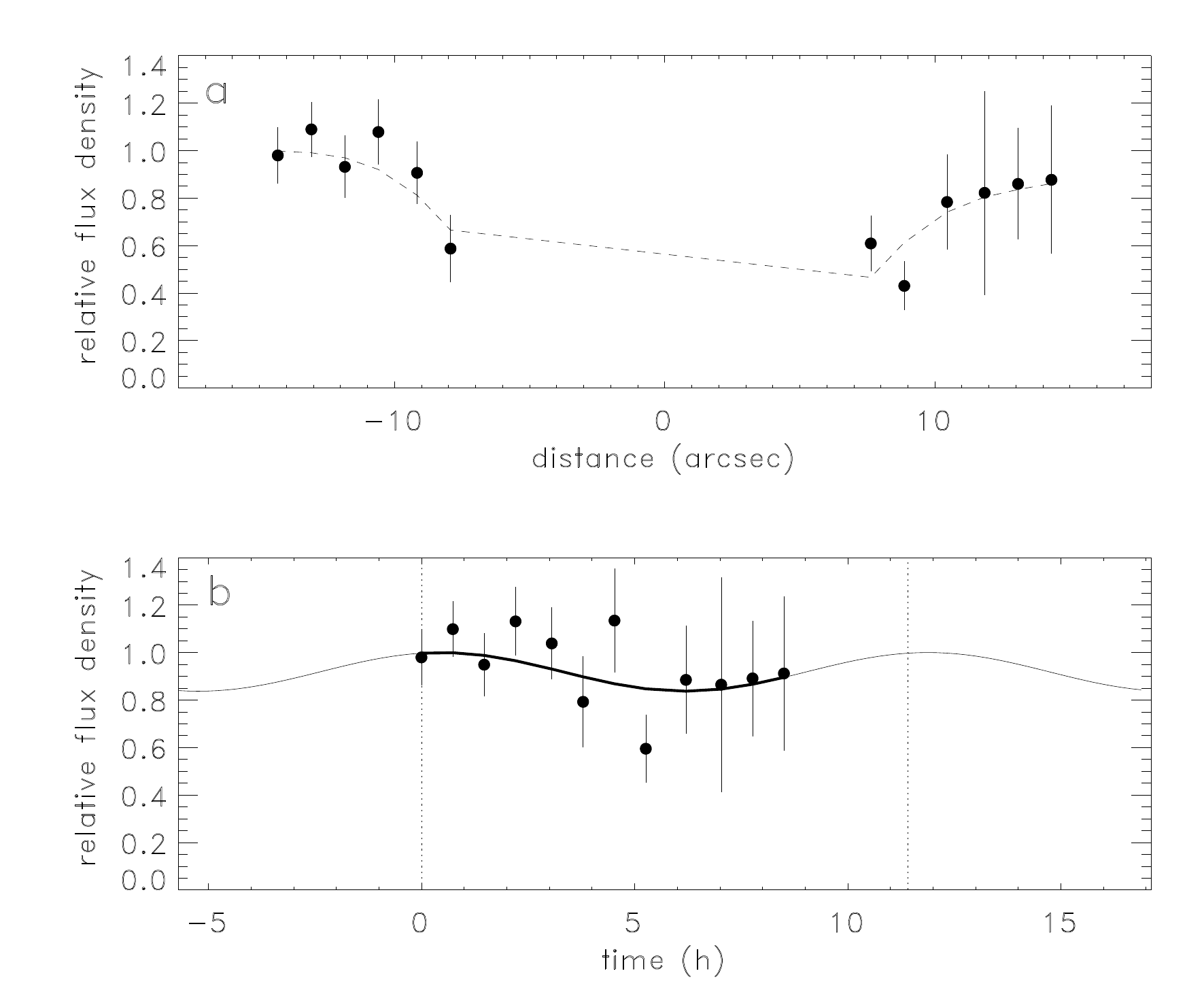}
    \caption{a) Relative flux densities as measured in the background-subtracted MIPS 24\,$\mu$m images, as a function of relative distance between the target and the mirror source. Negative and positive distances correspond to the first and second half of the measurement, respectively. The dashed curve represent the best-fit theoretical background-corrected relative flux density curve, considering flux loss due to the subtraction of the `mirror' source, and assuming a sinusoidal variation with a period of P\,=\,11.41\,h. 
    b) Variation of relative flux densities as a function of time (black dots) using the same best-fit sinusoidal fit as above (solid curve). The thicker part of the curve represent the phases covered by the MIPS 24\,$\mu$m observations. 
    }
    \label{fig:mipslc}
\end{figure}

This means that when the shadow images are subtracted, the 'mirror image' of Makemake will still affect the photometry on the background-subtracted images, notably decreasing the measured flux density. We simulated this effect by using the MIPS 24\,$\mu$m PSF and subtracted equal brightness sources at different distances from the target source. Using a small aperture with a radius of 3 pixels ($\sim$7\arcsec) the background subtraction has a very small effect at the largest, $\sim$14\arcsec\, distances, and we obtain $\geq$99\% of the true flux density; at the small distances ($\sim$8\arcsec), however, we obtain $\sim$60\% only. These simulated flux density ratios can be used to correct the measured background-subtracted flux densities. 

To check the possible flux density variations we used these simulated, mirror source distance dependent flux density corrections, also allowing for a sinusoidal 24\,$\mu$m light curve variation with a fixed period of P\,=\,11.4\,h, and we searched for the best-fit mean flux, amplitude and phase by residual minimization. The results are presented in Fig.~\ref{fig:mipslc}. The observed 24\,$\mu$m light curve can be best fitted by a low amplitude sinusoidal light curve with a relative peak-to-peak amplitude of 16$\pm$3\% or $\Delta F_{24}$\,=\,0.047$\pm$0.009\,mJy, over the $F_{24}^0$\,=\,0.295$\pm$0.025\,mJy mean in-band flux density. However, this light curve is also compatible with a constant light curve with 0.08\,mJy uncertainty ($\sim$25\% of the mean value). {We note that due to the uncertainties in the period determination the visible range light curves cannot be phased to the thermal emission measurements, as they are separated by several years.}

At 70\,$\mu$m we cannot perform the same background-subtraction as for the 24\,$\mu$m images due to the larger FWHM  of the PSF ($\sim$18\arcsec), i.e. the 70\,$\mu$m images may still contain contamination from background sources. We performed aperture photometry using an aperture radius of 8\arcsec\, and a background annulus between 39 and 65\arcsec\, on the pipeline processed (post-BCD) filtered (MFILT) images,  downloaded from the Spitzer Heritage Archive. We obtained a mean in-band flux density of $F_{71}$\,=\,11.17$\pm$0.50\,mJy for the AORKEYs 19176448...19179264, and the observed 70\,$\mu$m light curve is compatible with a constant light curve with a $\pm$2.93\,mJy uncertainty (25\% of the mean value). 

The same basic reduction steps were applied for the other measurement pair (AORKEYs 13803776/3264), without background subtraction (see Tables~\ref{table:allmeas} and \ref{table:all_flux}). The flux densities obtained in these cases are compatible with the average in-band flux densities obtained by \citet{Stansberry2008}: $F_{24}$\,=\,0.30$\pm$0.02\,mJy and $F_{71}$\,=\,14.6$\pm$2.2\,mJy.


\begin{table*}[ht!]
    \centering
    \begin{tabular}{ccccc|cc|c}
\hline
Instrument/ & OBSID /  & t$_{exp}$ & JD$_{mid}$ & band     & F$^i$ & data product & Ref.\\
    mode    & AORKEY  & (h)       &           & ($\mu$m) & (mJy) &  & \\
\hline
\hline
\multicolumn{8}{c}{James Webb} \\
\hline
MIRI       & V01254001001P0000000003101 & 0.14 & 2459975.0067 & 18.0 & 0.169$\pm$0.008 & & this work \\
                & V01254002001P0000000003101 & 0.14 & 2459975.0146 & 18.0 & 0.167$\pm$0.008 &    \\
                & V01254001001P0000000003103 & 0.14 & 2459975.0242 & 25.5 & 0.361$\pm$0.023 &    \\
                & V01254002001P0000000003103 & 0.14 & 2459975.0323 & 25.5 & 0.352$\pm$0.022 &     \\  
                & V01254001001P0000000002101 & 0.003 & 2459975.07637 & 5.6 & 0.058$\pm$0.003 & TAQ image\\
                \hline \hline
\multicolumn{8}{c}{Herschel} \\
\hline
SPIRE  & 1342187319,320,524,525    &  2.44 & 2455261.7851 & 250   & 11.8$\pm$1.2 &  & L17 \\
                & 1342198251,451            &   &   &       &       &         &     \\
SPIRE  & 1342187319,320,524,525    &  2.44 & 2455261.7851  & 350   & 9.0$\pm$1.2  &  & L17 \\
                & 1342198251,451            &   &   &       &       &         &     \\
SPIRE  & 1342187319,320,524,525    &  2.44 &  2455261.7851 & 500   & 4.4$\pm$1.4  &  & L17 \\
                & 1342198251,451            &   &   &       &       &         &     \\ \hline
PACS   & 1342187366 &   0.44 & 2455166.2915 &  70    &       12.06$\pm$1.54 & SPG/HPF & this work   \\
chop-nod        & 1342187367 &   0.44 & 2455166.3107 & 100   &       13.84$\pm$1.53 &  &   \\
                & 1342187366-367  &  0.88 & 2455166.3011 & 160 & 16.91$\pm$1.57       &                    &  \\ \hline
PACS   & 1342197657-7658  & 0.98 & 2455350.4194  & 70   &  11.57$\pm$1.08 & DIFF1        & this work    \\
scan maps       & 1342197695-7696  & 0.98 & 2455351.2381  & 70   &  9.36$\pm$1.08 & DIFF2        &     \\
                & 1342197659-7660  & 0.98 & 2455350.4406  & 100  &  15.62$\pm$1.22 & DIFF1        & this work    \\
                & 1342197697-7698  & 0.98 & 2455351.2592  & 100  &  19.58$\pm$1.22 & DIFF2        &     \\
                & 1342197657-7660  & 1.96 & 2455350.4300  & 160  & 22.09$\pm$2.76 & DIFF1        & this work    \\
                & 1342197695-7698  & 1.96 & 2455351.2485  & 160  & 17.99$\pm$2.76 & DIFF2        &     \\
\hline
\hline 
\multicolumn{8}{c}{Spitzer} \\
\hline
MIPS    & 13803776 &  0.41  & 2453542.2035 & 24    &  0.323$\pm$0.049 & BCD/MAIC & this work  \\
                & 13803264 &  0.41 & 2453543.4306 & 24    & 0.384$\pm$0.095  & &  \\ 
                & 13803776 &  0.41  & 2453542.2035 & 70    & 13.48$\pm$3.47 & BCD/MFILT & this work  \\
                & 13803264 &  0.41 & 2453543.4306 & 70    & 15.12$\pm$2.79  & &  \\  \hline
MIPS &	19179264-6448 &  7.584  & 2454257.1740 & 24 & 0.295$\pm$0.025$^*$  & BCD/MAIC  & this work	\\
             & 19179264-6448  & 7.584   & 2454257.1740 & 70 & 11.17$\pm$0.50$^*$   & BCD/MFILT & this work \\ 
\hline
\hline 
\multicolumn{8}{c}{ALMA} \\
\hline
ALMA            &               & 0.31    & 2457449.79514  & 1300  & 1.185$\pm$0.085    &    & L17 \\
\hline
\end{tabular}
\caption{Far-infrared and sub-mm observations of Makemake. The columns list the instruments, AORKEYs/OBSIDs, the mid times (Julian date), in-band flux densities (F$^i$), data product type, and the source of the photometric data. The SPIRE colour correction factors are 0.945, 0.948 and 0.943 at 250, 350 and 500\,$\mu$m, assuming a black body SED in the Rayleigh-Jeans regime. Summary of flux densities of Makemake obtained in previous evaluations and in this work. L17 is \citet{Lellouch2017}; measurements marked by $*$ are mean values of the long Spitzer measurements.    
} 
\label{table:all_flux}
\end{table*}
\paragraph{Herschel Space Observatory measurements}

Makemake was observed with the PACS photometer in chop-nod mode in the Science Demonstration Phase (SDP) \citep{Lim2010}. 
Note that this mode was used for faint source photometry early in the mission, and was replaced by the recommended (mini) scan map mode after the SDP \citep{Nielbock2013,Kiss2014}. Sources observed in chop-nod mode could have a serious contamination by background sources, as the final background of the image was made undecryptable by multiply folding the original 'chopped' and 'nodded' sub-images. \cite{Lim2010} obtained 11.4$\pm$2.7\,mJy, 12.0$\pm$2.8 and 16.7$\pm$3.5\,mJy at 70, 100 and 160\,$\mu$m. We re-evaluated this dataset, consisting of 70/160\,$\mu$m and 100/160\,$\mu$m filter combination measurements. We used the latest Herschel Science Archive Standard Product Generation level-2 high-pass-filtered images and performed the photometry in HIPE (version 15.0.0). We used the {\sl annularSkyAperturePhotometry()} task and the recommended apertures of 12\arcsec, 12\arcsec\, and 20\arcsec\, radii \citep{Nielbock2013} in the three bands, respectively, to obtain the flux densities,  and we used {\sl photApertureCorrectionPointSource()} to perform the encircled energy fraction correction \citep{Balog2014}. Flux density uncertainties are obtained using a set of circular apertures placed at a distance of 20\arcsec\, at 70 and 100\,$\mu$m and of 28\arcsec\, at 160\,$\mu$m \citep[see e.g.][]{Klaas2018}. The flux densities obtained are 12.06$\pm$1.54, 13.84$\pm$1.53 and 16.91$\pm$1.57\,mJy at 70, 100 and 160$\mu$m (also listed in Table~\ref{table:all_flux}). These flux density values are somewhat larger than those obtained by \cite{Lim2010} at 70 and 100\,$\mu$m, and very similar at 160\,$\mu$m. 

Makemake was also observed in scan map mode, in the framework of the `TNOs are Cool!' Open Time Key Program \citep{Muller2009}. The measurements followed the standard `TNOs are Cool!' measurements sequence, i.e. the target was observed at two epochs (referred to as Visit-1 and Visit-2), and the time between the two visits was set in a way that the target moved $\sim$30\arcsec{} with respect to the sky background that allowed us to use observations at the two epochs as mutual backgrounds. Observations at a specific visit also included scan/cross-scan observations in the same band, and in both possible PACS photometer filter combinations (70/160 and 100/160\,$\mu$m, see Table~\ref{table:allmeas}). Data reduction of the data is performed with the faint moving target optimized version of the PACS reduction pipeline \citep{Kiss2014} that produces Level-2 high-pass filtered maps from the individual scan map observations (OBSIDs). The scan and cross-scan images of the PACS band are combined to produce the `co-added' images for each epoch. The co-added images of the two epochs are further combined to obtain background-eliminated differential images. The `background matching' method {of} \cite{Kiss2014} is applied to correct for the small offsets in the coordinate frames of the Visit-1 and Visit-2 images using images of systematically shifted coordinate frames and then determining the offset that provides the smallest standard deviation of the per-pixel flux distribution in a pre-defined coverage interval -- the optimal offset can be most readily determined using the 160\,$\mu$m images due to the strong sky background w.r.t. the instrument noise. The differential (DIFF) images contain a `positive' and a `negative' source separated by $\sim$30\arcsec, corresponding to the two observational epochs. While higher level data products also exist \citep[e.g double-differential images, see][]{Kiss2014} in this paper we used the DIFF images to obtain photometry, separately at the two epochs. Flux densities are obtained using the faint source optimized `TNOs are Cool!' photomery pipeline, including the derivation of the flux density uncertainties using the implanted source method \citep{Kiss2014}. The results are presented in Table~\ref{table:all_flux}. 

The Herschel/SPIRE sub-mm photometer also observed Makemake, the latest re-evaluation of the results are available in \citet{Lellouch2017}. The monochromatic flux densities obtained are 12.5$\pm$1.3\,mJy, 9.5$\pm$1.3\,mJy, and 4.7$\pm$1.5\,mJy at 250, 350 and 500\,$\mu$m, respectively, from multi-epoch observations. We used these combined flux densities in our subsequent analysis.

\paragraph{James Webb Space Telescope MIRI imaging observations}

Makemake was observed in the framework of the GTO program 1254 `TNOs' (P.I. A.~Parker) in the F1800W and F2550W bands of the MIRI imaging instrument \citep{MIRI} of the James Webb Space Telescope.  
{The JWST data presented in this article were obtained from the Mikulski Archive for Space Telescopes (MAST) at the Space Telescope Science Institute. The specific observations analyzed can be accessed via \dataset[10.17909/0ebf-dp91]{https://doi.org/10.17909/0ebf-dp91}}. 
To ensure the most up-to-date calibration, the raw  {\sl uncal} files were downloaded from the Mikulski Archive for Space Telescopes (MAST) and processed locally using version 1.15.1 of the JWST calibration pipeline \citep{2023zndo...6984365B} and the {\sl jwst\_1256.pmap} reference file context. None of the default pipeline parameters were changed when processing the science, background, or target acquisition observations. The fully calibrated, non-distortion corrected {\sl cal} files were used in this investigation. In addition to the MIRI photometric measurements, Makemake was identified on the target acquisition image of the MIRI/{LRS} measurement V01254001001P0000000002101, obtained with the F560W filter. Aperture photometry provided an aperture-corrected flux density of 58\,$\pm$3\,$\mu$Jy.

\section{Possible contaminating sources and colour correction effects \label{sect:contamination}}

As Makemake seems to be significantly brighter than expected, especially in the JWST MIRI F1800W filter, we investigated the probability that the JWST measurements may be contaminated by another source. First we checked the positional uncertainty of Makemake's orbit. According to JPL-Horizons, the astrometric 3$\sigma$ uncertainties for the date of the JWST/MIRI observations are $\sigma_{RA}$\,=\,22\,mas, and $\sigma_{DEC}$\,=\,15\,mas, in right ascension and declination, respectively (i.e. Makemake is expected to be within these limits with 99.7\% probability). Indeed, centroid fitting provides astrometry within $\lesssim$100\,mas in all images to the predicted positions, which is much smaller than the $\sim$591\,mas beam FWHM of the F1800W filter. To estimate the probability that an extragalactic source with the observed flux density is located within these uncertainties, we used the number count predictions by \citet{Cowley2018}. For the F1800W filter the expected number of sources $N({>}\,150\,\mu Jy)$\,=\,0.5\,arcmin$^{-2}$, and the corresponding probability that such a source is within the astrometric uncertainty is p\,=\,1.1$\times$10$^{-6}$. Also, while Makemake moved $\sim$1\arcsec\, between the first F1800W and the last F2550W measurements (see Table~\ref{table:allmeas}), the fitted astrometric positions of the centroids remained consistent with the predicted positions of the target, i.e. we can safely reject the assumption that the source is an extragalactic, or other sidereal source. 

Two known asteroids ((400986) and 2021\,KV$_{111}$) were identified in the relative vicinity of Makemake at the time of the JWST/MIRI observations, but both asteroids were at safe distance ($>$30\arcsec) from Makemake. Due to their higher surface temperature, small (undiscovered) main belt asteroids in the few hundred meter size range may have F1800W and F2550W flux density values similar to the observed ones \citep[see][for a serendipitous detection of a similar main belt asteroid with JWST]{Muller2023}. However, typical main belt asteroids move with a speed of $\sim$0.01\arcsec\,s$^{-1}$, which is clearly inconsistent with the observed movement. While a Centaur or a trans-Neptunian object could mimic the apparent motion of Makemake, it should be extremely large (at the top of the observed size range of these objects) to produce the observed flux densities. Therefore we can safely consider the source observed with JWST/MIRI to be Makemake itself. 

Due to the general cold temperatures that are expected to be associated with the surface of Makemake, short wavelength mid-infrared filters in our analysis may experience large colour correction factors.  Colour corrections perform the transformation between the in-band and monochromatic flux densities: $F_m(\lambda)$\,=\,$F_i(\lambda)/C(\lambda)$, where $F_m(\lambda)$\, and \,$F_i(\lambda)$ are the monochromatic and in-band flux densities, respectively, and $C(\lambda)$ is the wavelength and spectral energy distribution dependent colour correction factor, which accounts for the effect of a spectral energy distribution which is different from the reference spectral energy distribution used in the photometric calibration of the instrument \citep[see e.g.][]{Gordon2022}. 

Based on the visible range geometric albedo, the associated phase integral, and the heliocentric distance, Makemake is expected to have dayside average surface temperatures of $\sim$40\,K. In the case of a short wavelength (mid-infrared) excess with respect to that of these cold terrains, the mid-infrared detectors will see higher effective temperatures. 
The colour correction factors of the mid-infrared filters are relatively small ($C(\lambda)$\,$\lesssim$\,1.5), although not negligible, for Makemake's typical surface temperatures (T\,$\approx$40\,K), associated with a single-terrain model (see Sect.~\ref{sect:thermalmodelmain}). For higher temperatures $C(\lambda)$\,$\approx$\,1, i.e. colour corrections do not affect the mid-infrared flux densities notably. 


These calculations assumed a pure black body thermal emission. If there are no other sources but the cold surface of Makemake that contributes to the thermal emission, there is a considerable contribution from the reflected light of Makemake's surface to the mid-IR SED, especially at 18\,$\mu$m (see Sect.~\ref{sect:thermalmodelmain}). This makes the mid-infrared JWST/MIRI F1800W, F2550W and the Spitzer/MIPS 24\,$\mu$m colour correction factors much smaller than that of a single black body. In our models discussed above all colour correction factors remain $C(\lambda)$\,$\lesssim$\,1.1, i.e. they do not change considerably the monochromatic flux densities, and we can safely state that the high mid-infrared flux densities cannot be explained by colour correction effects. 
In the thermal emission modeling (Sect.~\ref{sect:thermalmodelmain}) colour correction were considered for all filters, using the colour correction factors calculated from the actual (model) SEDs and the filter transmission curves.  

To our knowledge, no filter leak has been identified for the MIRI F1800W or F2550W filters \citep[see][for the evaluation of the MIRI imaging in-flight performance]{Dicken2024}. 


\section{Thermal emission modeling \label{sect:neatmmod}}

The shorter wavelength mid-infrared (5\,$\leq$\,$\lambda$\,$\leq$\,10\,$\mu$m) light from Makemake is expected to be dominated by reflected light. The expected flux densities of the reflected light from Makemake were calculated for the observing geometry of the JWST/MIRI measurements using two methods. We first considered a D\,=\,1430\,km sphere at Makemake's heliocentric and observer distance and phase angle, and using the NASA SSI solar reference spectra\footnote{https://sunclimate.gsfc.nasa.gov/ssi-reference-spectra}. We also calculated it from the known absolute brightness and observing geometry, with a wavelength scaling that corresponds to a 5780\,K black body. The two calculations resulted in nearly identical reflected light SEDs.
The JWST/MIRI {LRS} target acquisition measurement provided an F560W band flux density of F$_{5.6}$\,=\,58$\pm$3\,$\mu$Jy which, when scaled to the solar reference spectrum, corresponds to a $p_{5.6}$\,=0.80$\pm$0.05 albedo at 5.6\,$\mu$m. We used this albedo value as representative for all $\lambda$\,$\geq$\,5\,$\mu$m wavelengths when considering the reflected light contribution in thermal emission calculations.

To model the thermal emission of Makemake we use an elliptical NEATM model \citep{Farkas2020}, however, the potentially non-spherical shape of Makemake plays little role. As it has been previously suggested by \citet{Ortiz2012} and \citet{Brown2013}, the apparent axis ratio of Makemake, as obtained from the occultation chords, is $a'/b'$\,$\lesssim$\,1.05. Also, using the mass of Makemake \citep{Parker2018}, the occultation size and the rotation period of P\,=\,11.4\,h, a Maclaurin spheroid -- the shape of a rotationally deformed strengthless body held together by gravity --  would have a true axis ratio of $a/c$\,$\approx$\,1.05, which translates into the same apparent axis ratio ($a'/b'$\,$\approx$\,1.05) as in the case of an equator-on geometry. Using this axis ratio, the largest possible apparent semi axes are $a'$\,=\,751\,km and $b'$\,=\,715\,km. This leads to a spectral energy distribution which is only slightly, $\sim$2-4\% different from the spherical case of $a'$\,=\,$b'$\,=\,715\,km, as obtained from the occultation results. Therefore in the following we consider a spherical Makemake in the thermal emission modeling. 

Some of the scenarios we investigate here have been considered in \citet{Parker2016} and \citet{Lellouch2017}, but here we use additional thermal emission data, updated flux densities, and different ancillary data (e.g. new phase integral values) than in these previous papers. 

\paragraph{Case 1: Single terrain Makemake \label{sect:singleterrain}}
First we considered Makemake to have a single-terrain surface with a geometric albedo of $p_V$\,=\,0.82$\pm$0.02, phase integral of $q$\,=\,0.90$\pm$0.05, and therefore Bond albedo of A$_B$\,=\,0.74$\pm$0.06 \citep{V22}. We assume that the surface is homogeneous, and no other component in the system contributes to the thermal emission. We calculated NEATM models using beaming parameters of $\eta$\,=\,0.6--2.6 \citep{Lellouch2013}, presented as gray curves in Fig.~\ref{fig:neatm3}a.  
The thermal emission model of $\eta$\,=\,1.2, the mean value among trans-Neptunian objects \citep{Stansberry2008,Vilenius2012,Lellouch2013} fits the observed flux densities very well for $\lambda$\,$\geq$\,160$\mu$m. 


\paragraph{Case 2: Single terrain Makemake + dark satellite \label{sect:neamt_sat}} 
Makemake has a satellite which may contribute significantly to the thermal emission in the mid-infrared and at the shorter far-infrared wavelengths, as discussed previously by \citet{Parker2016} and \citet{Lellouch2017}. 
The satellite is 7.8\,mag fainter than Makemake \citep{Parker2016}. The absolute magnitude of the Makemake system is H$_V$\,=\,0.049$\pm$0.020\,mag \citep{Hromakina2019}, i.e. the satellite's absolute magnitude is H$_V$\,=\,7.85\,mag. This would correspond to diameters D\,=\,51, 113, 160 and 253\,km assuming geometric albedos of p$_V$\,=\,0.5, 0.1, 0.05, and 0.02, respectively. The satellite contribution is a very small correction on Makemake's absolute brightness, Makemake alone has H$_V$\,=\,0.050\,mag. The phase integral for Makemake is q\,=\,0.90$\pm$0.05, and for a dark moon (p$_V$\,$\leq$\,0.05) a typical phase integral in the trans-Neptunian region is q\,=\,0.26, as determined by \citet{V22}. We again used the NEATM model, as in the single terrain case, assuming the same model for the bright terrain. In addition, we considered the thermal emission of the satellite assuming $p_V$\,=\,0.01, 0.02, 0.04 and 0.08, and using a beaming parameter of $\eta$\,=\,0.6 in all these cases. $\eta$\,=\,0.6 is about the physically possible minimum value for the beaming parameter \citep{Lellouch2013}, and provides the highest surface temperatures (highest mid-infrared flux density) for the same albedo. 
In addition, we also included the 'extreme moon' model by \citet{Lellouch2017} which assumed $p_V$\,=\,0.017 and $\eta$\,=\,0.34. The results as presented in Fig.~\ref{fig:neatm3}b. 

\paragraph{Case 3: Double terrain Makemake \label{sect:neatm_double}}

To test whether the warm, short wavelength emission in the Makemake system could be explained by a dark terrain on the surface of Makemake, as suggested e.g. by \citet{Lim2010}, we also calculate double terrain models. For all double terrain configurations we assume that the bright terrain is Eris-like ($p_{V1}$), and the dark terrain ($p_{V2}$) is either Charon-like or Quaoar-like. The relative contribution of these two terrains can be obtained according to the equation, neglecting the potential differences in scattering properies of the two terrains, $p_{V0} = \Omega_1 p_{V1} + \Omega_2 p_{V2} $, 
where $p_{V0}$ is the global (observed, disk-integrated) geometric albedo of Makemake, $p_{V1}$ and $p_{V2}$ are the geometric albedos of the two terrains, and $\Omega_1$ and $\Omega_2$ are the relative areas of the two terrains ($\Omega_1$+$\Omega_2$\,=\,1). Following \citet{Ortiz2012} we assume that the dark terrain is located around the subsolar latitude belt, which depends on the orientation of the axis of rotation. As discussed in \citet{Brown2013} previous thermal emission constraints and the very small light curve amplitude indicate that Makemake might be seen very close to a pole-on configuration, with an aspect angle of $\vartheta$\,$\approx$\,20\degr. However, Makemake's satellite \citep{Parker2016} 
passes very close to Makemake in the line of sight, indicating that the satellite orbit may be close to edge-on, also suggesting an edge-on rotational configuration in the case of a tidally evolved system. 
The properties of the terrains we use are summarized in Table~\ref{table:terrains} below. 

\begin{table}[ht!]
    \centering
    \begin{tabular}{cccccc}
    \hline
     \multicolumn{2}{c}{Terrain}  & p$_V$ & q & A$_B$(=A$_V$) & A$_B'$ \\
    \hline 
     Makemake & single &0.80 & 0.90 & 0.74 & 0.74 \\
     \hline
     Eris     & bright & 0.96 & 1.04 & 1.00 & 0.90 \\
     Charon   & dark & 0.42 & 0.60 & 0.25 & 0.25 \\
     Quaoar   & dark & 0.11 & 0.52 & 0.057 & 0.057 \\
    \hline
    \end{tabular}
    \caption{Properties of the terrains considered for the NEATM thermal emission modeling of Makemake. The columns are: 1) name of the terrain, corresponding to the global properties of the bodies named. 2) V-band geometric albedo; 3) V-band phase integral; 4) Bond albedo obtained from visible band measurements; 5) Bond albedo used for thermal emission modeling. All values are obtained from \citet{V22} except the A$_B'$ value of Eris, which is obtained from spectral modeling (see the text for details).   }
    \label{table:terrains}
\end{table}

A$_B'$ of the Eris terrain is calculated using a synthetic reflectance spectrum of Eris, obtained by assuming pure CH$_4$ grains with a specific size distribution, as obtained by \cite{AC2020}. Bond albedo is calulated using a version of the \citet{Hapke} reflectance model, used for modeling the low phase angle surface of Eris \citep{Trujillo2005,SZK23}. This wavelength-dependent physical albedo is convolved with the solar irradiance spectrum to obtain the total power per surface area (W\,m$^{-2}$) reflected from the surface in the visible and near-infrared range (0.3--3\,$\mu$m). This provides $A_B'$\,=\,0.90, clearly showing that the usual assumption of A$_B$\,$\approx$\,$A_V$ in asteroid thermal emission models does not hold in the case of the surface of Eris, which is highly reflective in the visible. For thermal emission modelling, the fraction of the power of solar irradiation absorbed is (1--$A_B'$), and we use this value for the Eris-like terrain. For all other terrains we assume that the usual A$_B$\,$\approx$\,$A_V$ holds, and we use the $A_B$ values obtained by \cite{V22} in the thermal emission modeling, too.  

As shown in Fig.~\ref{fig:neatm3}c, all dark terrain types discussed above show very similar spectral energy distributions (green, yellow and red stripes) if the same beaming parameter ($\eta$) is used. Models using terrains with $\eta$\,=\,1.0 (or higher) underestimates the mid-infrared flux densities while being roughly compatible with the long-wavelength ($\lambda$\,$\geq$\,70\,$\mu$m) flux densities. 


\section{Dust temperature \label{sect:dusttemperature}}

\begin{figure}[ht!]
    \centering
    \includegraphics[width=0.96\columnwidth]{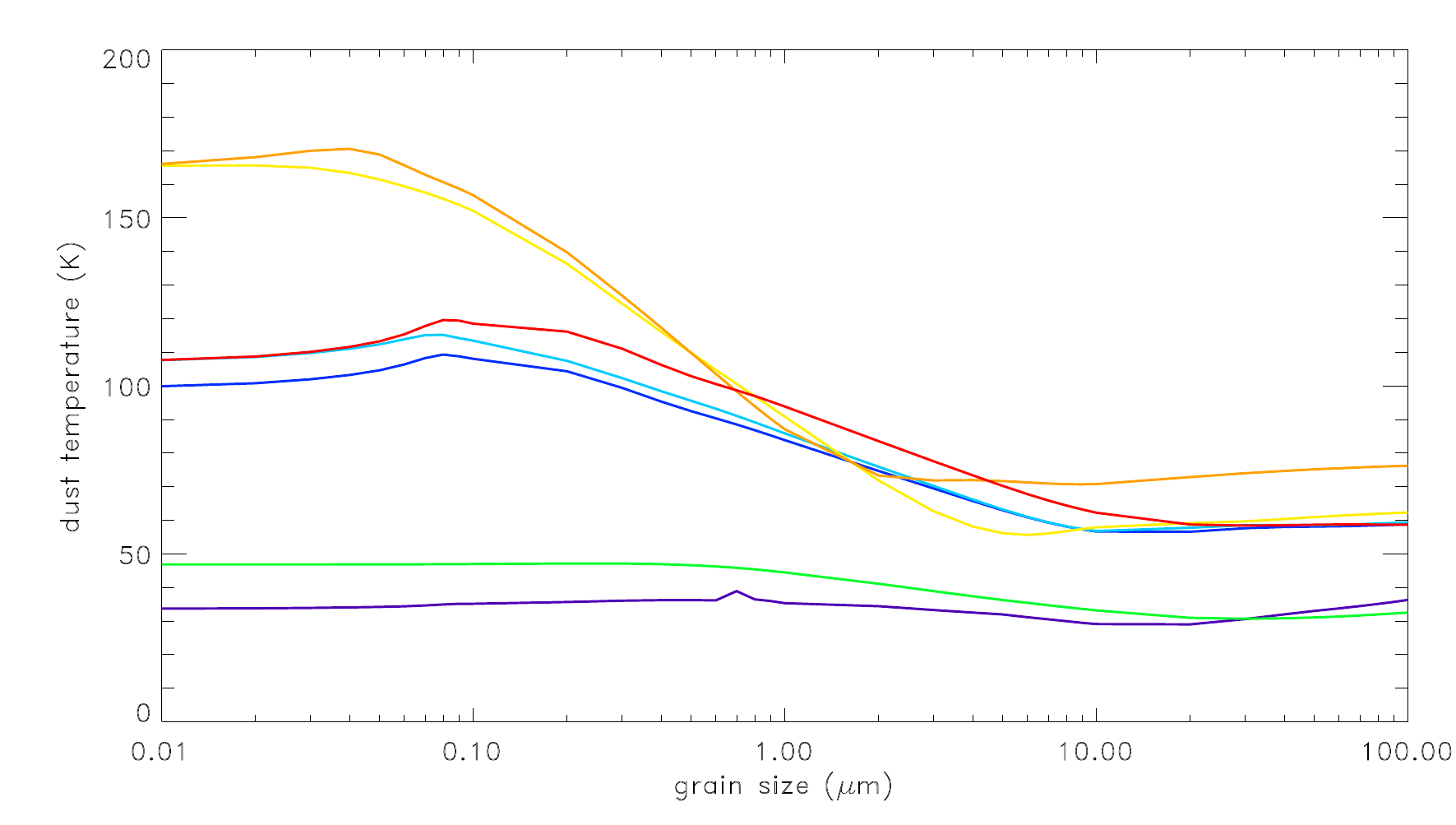}
    \caption{Dust temperature as a function of grain size for different compositions. Purple: pyroxene with 100\% Mg-content,  blue: pyroxene with 40\% Mg-content, light blue: olivine with 50\% Mg-content, green: amorphous water ice, yellow: carbon grain as defined by \citet{Zubko1996}, orange: graphite grains, red: 'organic carbon'.}
    \label{fig:dusttemp}
\end{figure}

We calculated the possible equilibrium dust temperatures at the heliocentric distance of Makemake for a range of possible materials and grain sizes, obtained from the power balance between the absorbed solar radiation and thermal emission of the grains, using the equation:
\begin{equation}
    \frac{R_\odot^2}{r_h^2}\int_0^\infty B_\lambda(\lambda,T_\odot)Q_{abs}(\lambda)d\lambda = 
    \int_0^\infty B_\lambda(\lambda,T_d)Q_{abs}(\lambda)d\lambda
\end{equation}
where $R_\odot$ is the radius of the Sun, $r_h$ is the heliocentric distance, $T_\odot$ the temperature of the Sun's photosphere, $Q_{abs}$ the absorption coefficients of the grains, and $T_d$ the dust temperature. 
Absorption coefficients have been determined using {\sl OpTool} \citep{optool}, for each specific single grain size and composition. We have considered different kinds of carbonaceous and silicate grains, as well as water ice. The estimated dust temperatures  are presented in Fig.~\ref{fig:dusttemp}. 

\section{Radiative transfer modeling of the ring \label{sect:ringrad}}

Radiative transfer modeling of the dust grains in the ring is performed using RADMC-3D \citep{radmc}. 
The materials used for radiative transfer modeling are characterized by their composition, porosity and size distribution. We use the {\sl OpTool}  package \citep{optool} to obtain the absorption and scattering coefficients of the grains, using materials from {\sl OpTool}'s material library, and assuming a specific grain size distribution. We consider different types of materials in the different simulations, but we assume that the whole simulation volume is homogeneous material-wise. Anisotropic scattering is considered, and treated by applying the Henyey-Greenstein function (option {\sl scattering\_mode\_max\,=\,2} in RADMC-3D), and using the scattering opacity and $g$ anisotropy parameters obtained for the specific material with {\sl OpTool}.

We assume that an approximate ring model can be calculated by modeling the radiative transfer in a small element of the ring, and simply multiplying the flux densities by the ratio of the ring area ($A_r$) to the area of the simulated region ($A_s$).  
The apparent area of the ring is: 
\begin{equation}
    A_r = \pi( (r_i+w)^2 - r_i^2 )\sin B
\end{equation}
where $r_i$ and $w$ are the inner radius and width of the ring, and B is the opening angle. 
This model implicitly assumes that the ring is homogeneous, thin ($d\,\ll\,w$, where d is the thickness of the ring), and that edge effects are ignored. 
In detail we used a 128$\times$128$\times$8 cell simulation volume, which is illuminated under the proper solar geometry with respect to the ring, and seen under the proper geometry from Earth (the smallest dimension corresponds to the vertical direction, perpendicular to the ring plane). The simulation volume is filled homogeneously with the actual material chosen. 
\begin{figure}[ht!]
    \centering
    \includegraphics[width=\columnwidth]{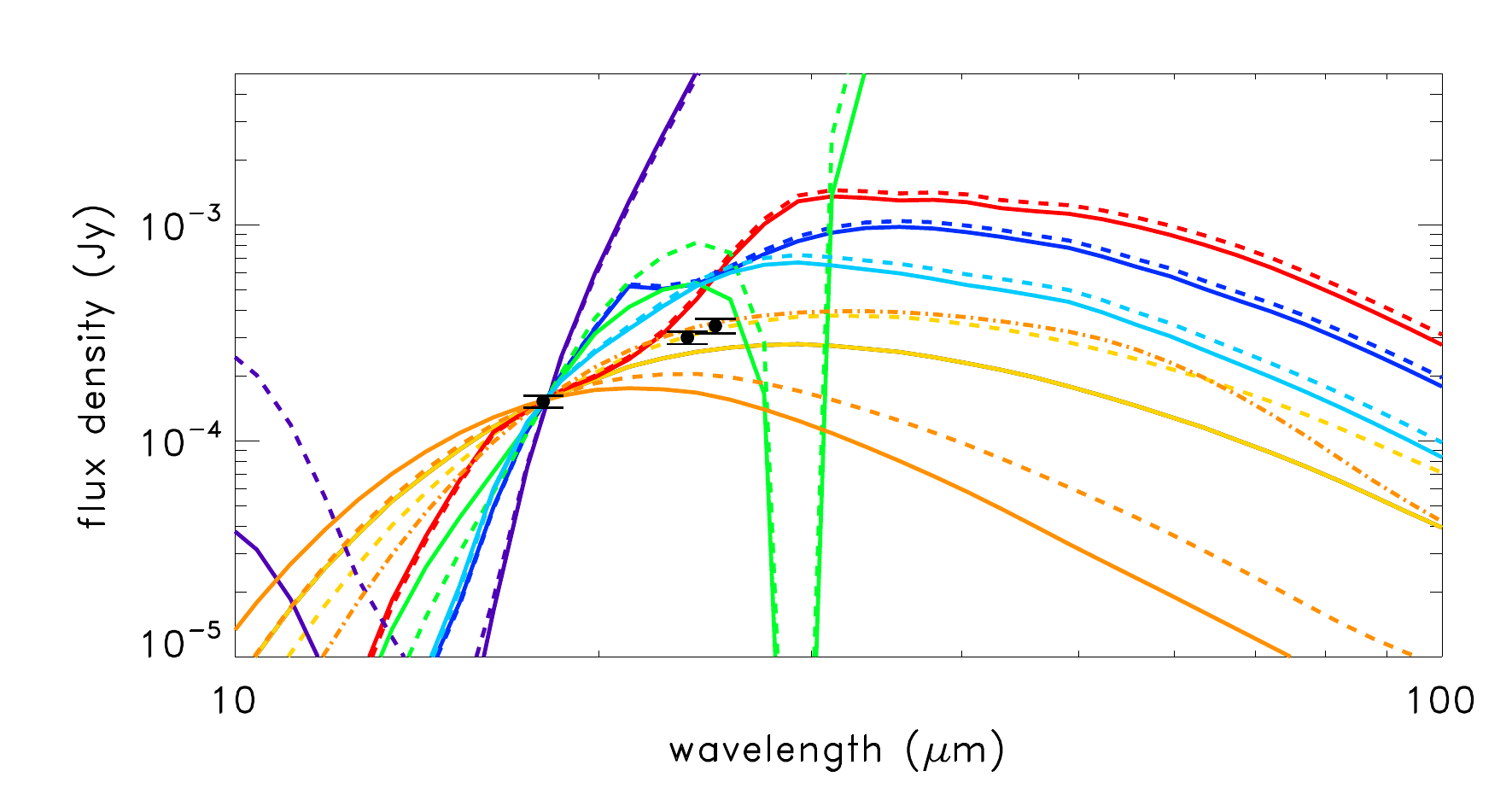}
    \caption{Spectral energy distribution of different potential ring materials, normalized to the observed flux density at 18\,$\mu$m (JWST/MIRI F1800W filter). The colors are the same as in Fig.~\ref{fig:dusttemp}. Solid and dashed curves correspond to the same material, but using 100\,nm and 200\,nm grain sizes, respectively. The orange dash-dotted curve is for graphite grains with 500\,nm grain size. The black symbols with error bars are the mean MIRI F1800W, F2550W, and MIPS 24\,$\mu$m flux densities. }
    \label{fig:dustseds}
\end{figure}
\begin{figure*}
    \centering
    \includegraphics[width=0.8\textwidth]{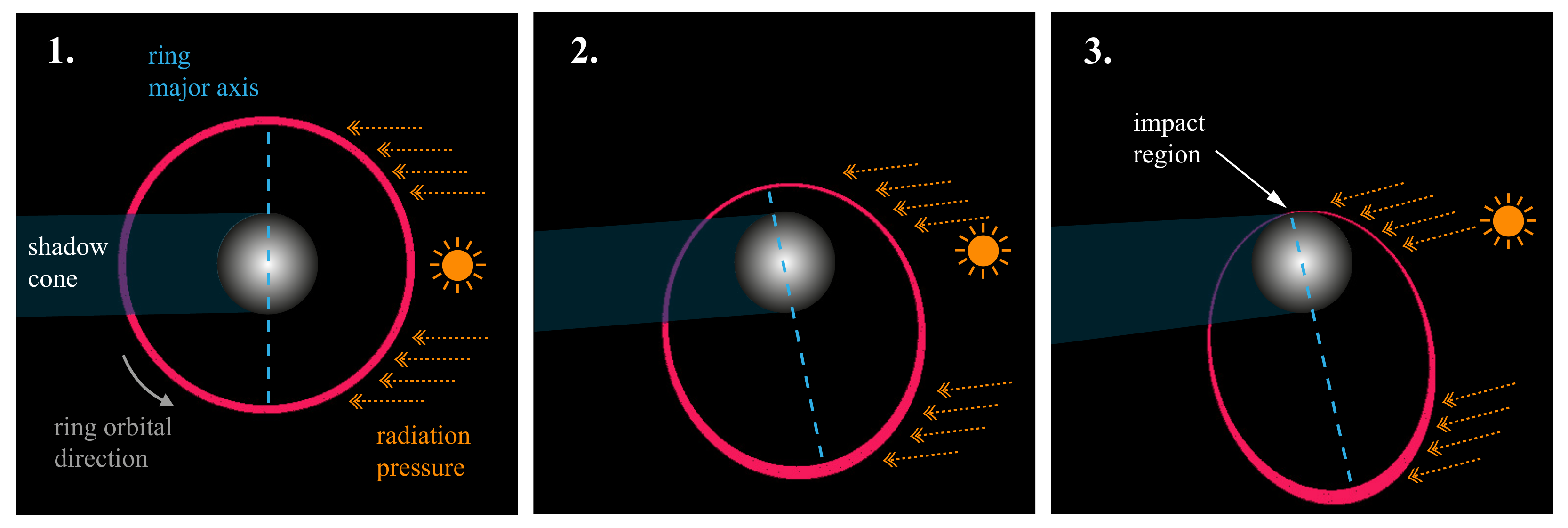}
    \caption{Time evolution (from left to right) of small grains around Makemake under the influence of solar radiation pressure. }
    \label{fig:eccevo}
\end{figure*}

The apparent, projected image of the simulated ring element is not homogeneous due to projection effects, and has lower intensity / optical depth towards the edges. However, it has the correct values in its central part, which corresponds to the surface brightness values as if it were being seen in a homogeneous, full ring. We used those projected pixels for which $\tau$\,=\,max($\tau$) in the visible range R-band, also calculated by RADMC-3D.
We use this model to calculate both the reflected light and the thermal emission components.  We assume that the ring is seen under an opening angle of B\,=\,15\degr, and has a width of $w$\,=\,10\,km. Until the ring is optically thin the ring width has a negligible effect on the final SED, but the observing and illumination geometry, including the B opening angle, is important, especially when calculating the reflected light, due to the directional dependence of the scattering. 
Example spectral energy distributions of the ring, assuming different materials, are shown in Fig.~\ref{fig:dustseds}. 

\section{Grain removal timescales \label{sect:grainremoval}}

We used the approach by \citet{Burns1979} to compute the evolution of dust grains around Makemake using a dynamical model, considering the gravitational field of Makemake and the Sun, and the solar radiation pressure. 
The radiation pressure force component is obtained as:
\begin{equation}
    {F_r} = {S_0 (r_0/r_h)^2 \pi s^2 c^{-1} Q_{pr}}
\end{equation}
where $S_0$ is the solar constant, $r_0$ is 1\,au, $r_h$ is the heliocentric distance of Makemake in [au], $s$ is the grain radius, $c$ the speed of light, and $Q_{pr}$ the radiation pressure coefficient. $Q_{pr}$ is calculated using the optical constants of the grains, as obtained in the previous calculations, and it is grain size and composition dependent \citep[see][for a detailed discussion]{Burns1979}. 

\begin{figure}[ht!]
    \centering
    \includegraphics[width=0.99\columnwidth]{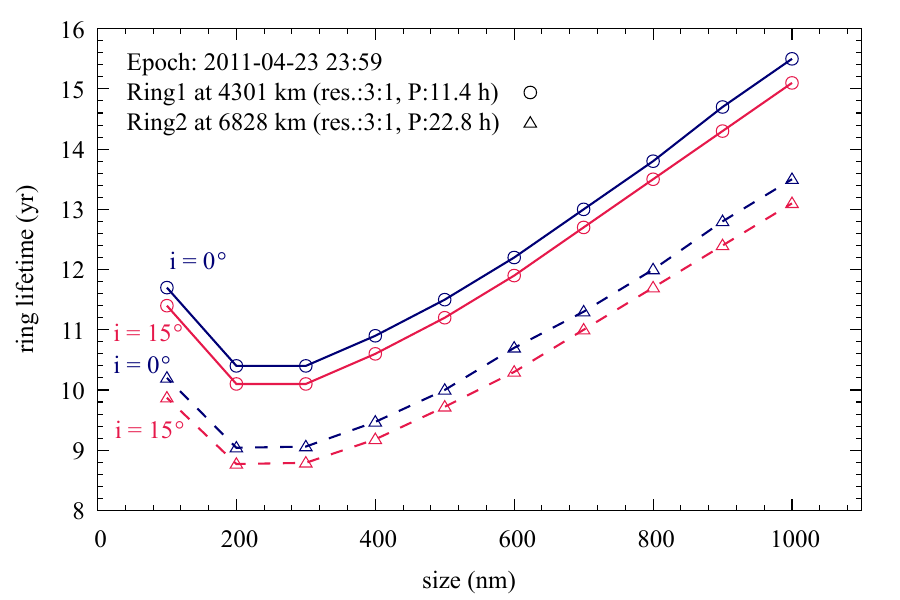}
    \caption{Lifetime of carbonaceous dust grains in the putative ring as a function of grain size. Curves with circle and triangle symbols mark the starting distances that correspond to the 3:1 spin-orbit resonances with P\,=\,11.4\,h and 22.8\,h, while red and blue curves correspond to ring opening angles of 0\degr\, and 15\degr, respectively. We used Makemake's heliocentric distance at the time of the occultation on April 23, 2011.}
    \label{fig:lifetime}
\end{figure}

The effect of radiation pressure is modelled using a high-precision GPU-assisted Hermite N-body integrator \citep[see][]{Nitadori2008,Regaly2018}
which takes into account the radiation pressure and the shadow cast on the ring by the dwarf planet. 
We performed the calculations for carbonaceous grains, for different grain sizes, for ring opening angles of B\,=\,0 and 15\degr, and for initial ring radii corresponding to the 3:1 spin orbit resonances assuming P\,=\,11.4 or 22.8\,h. In the initial phase, the ring particles are in circular orbits around Makemake. In the case of a non-negligible radiation pressure force, the photons radiated by the Sun perturb the particles' orbits, exciting their orbital eccentricity while maintaining their semi-major axes (see Fig.~\ref{fig:eccevo}). As time progresses, the ring becomes globally eccentric, with the pericentre and apocentre points situated at the positions where the radiation pressure accelerates and decelerates the ring particles, respectively. Due to the relative direction of the Sun changing as a result of the orbital motion of Makemake, the ring's major axis experiences continuous  tilting. In the end the ring's eccentricity reaches a critical value at which point $a_\mathrm{r}(1-e_\mathrm{crit})=R$, where $R$ is the radius of Makemake's body, i.e. the pericentre of the ring particles reaches the surface of Makemake, and consequently these grains are removed from the ring. 
As shown in Fig.~\ref{fig:lifetime}, small grains ($\leq$0.5\,$\mu$m) around Makemake have lifetimes of $\sim$10\,yr, with some moderate dependence on the starting conditions (ring opening angle and starting semi major axis). 

\end{document}